\documentclass[a4paper,11pt]{article}
\pdfoutput=1
\usepackage{jheppub}

\DeclareSymbolFont{usualmathcal}{OMS}{cmsy}{m}{n}
\DeclareSymbolFontAlphabet{\mathcal}{usualmathcal}

\DeclareMathOperator{\Tr}{Tr\,}
\DeclareMathOperator{\tr}{tr\,}

\usepackage{tensor}
\usepackage[most]{tcolorbox}
\usepackage{slashed}
\usepackage{comment}
\usepackage{tikz}
\usetikzlibrary{decorations.markings,arrows.meta}

\usepackage{xcolor}

\definecolor{GoldYellow}{HTML}{FFC107}

\newcommand{\p}{\partial}

\newcommand{\ppsi}{\chi}

\def\be{\begin{equation}} 
\def\ee{\end{equation}} 
\newcommand{\bal}{\begin{align}} 
\newcommand{\eal}{\end{align}}

\def\nl{\nonumber \\}

\def\a{\alpha} 
\def\b{\beta} 
\def\g{\gamma}

\def\d{\delta} 
 
\def\eps{\varepsilon}

\def\s{\sigma}

\def\th{\theta}

\title{ Hydrodynamics of perfect fluids with anomalies
    from the fermionic path integral }

\author[a]{Alexander G. Abanov}
\author[b,1]{Andrea Cappelli,\note{Corresponding author.}}

\affiliation[a]{ Department of Physics and Astronomy,
  Stony Brook University, Stony Brook, NY 11794, USA}
\affiliation[b]{ INFN, Sezione di Firenze,Via G. Sansone, 1,
  50019 Sesto Fiorentino,Firenze, Italy}

\emailAdd{alexandre.abanov@stonybrook.edu}
\emailAdd{andrea.cappelli@fi.infn.it}
  
\abstract{
The path integral of the Dirac fermion with vector and axial gauge
backgrounds is analyzed near the infrared limit in the presence of
residual irrelevant current-current interaction. After integrating out
fermions, a semiclassical low-energy effective action is obtained,
written in terms of currents.  Its expression is found to correspond
to the hydrodynamic action previously proposed for perfect barotropic
fluids with anomalies at zero temperature.  This approach also leads
to two further hydrodynamic actions to be associated, respectively,
with the Weyl fermion, and the Dirac fermion having independent vector
and axial currents.  These actions feature four- and five-dimensional
bulk-boundary terms, owing to anomaly inflow, which are identified as
being the so-called transgression forms.  These are generalizations of
Chern--Simons forms that involve two gauge fields: the dynamical field
and the background field.

The path-integral argument provides a ``microscopic'' explanation for
several ingredients of the action formulation of hydrodynamics that
are necessary to incorporate anomalies. It also clarifies the infrared
reduction required to pass from the effective field theory to a local
hydrodynamic description. This reduction is implemented by considering
restricted variations of the action, familiar from hydrodynamics,
which at the same time lead to four-dimensional equations of motion
from the five-dimensional transgression terms.}

\begin{document} 
\maketitle
\flushbottom

\section{Introduction}
\label{sec:intro}

The action formulation of hydrodynamics
\cite{schutz1970perfect,mobbs1982variational,carter1988standard} is
very convenient for establishing connections with effective field
theory \cite{Ratt2011,Liu:2018kfw,Jensen:2018hse,Rattfluid}.
The relation is useful in both directions. On the one hand, it
makes it possible to incorporate field-theoretic phenomena, such as
chiral anomalies
\cite{bertlmann2000anomalies,treiman2014current,arouca2022quantum},
directly into the hydrodynamic description
\cite{monteiro2015hydrodynamics,AbanovII,wiegmann2022hamilton,WiegWZ,
  abanov2024hydrodynamics,WiegCS}.
On the other hand,
hydrodynamics may suggest extensions of bosonization beyond two
spacetime dimensions
\cite{Fradkinloop,gaiottokapustinspinTQFT1,kapustinthorngren2017,
  CappelliVillaBosDual2025}. More
generally, the action approach makes the geometric structure of Euler
hydrodynamics \cite{lichnerowicz1994relativistic,carter1979perfect,NekWieg}
especially transparent and gives a unified description
of both Euclidean and Lorentzian fluids.\footnote{See e.g. Appendix A
  of \cite{abanov2024hydrodynamics}.}

In a previous paper \cite{abanov2024hydrodynamics}, we extended the
action formulation of barotropic Euler hydrodynamics to include chiral
anomalies in both two and four spacetime dimensions. In two
dimensions, we showed that the resulting theory is completely
equivalent to standard field-theoretic bosonization. The anomaly inflow
\cite{arouca2022quantum} from a $(d+1)$-dimensional topological theory
was used to identify the appropriate fluid degrees of freedom.

In four dimensions, the same
inflow argument suggested an extension of the hydrodynamic action by an
additional pseudo-scalar field, so as to properly describe
anomalies for both vector and axial gauge backgrounds.
We also included coupling to the gravitational field and the
corresponding mixed axial-gravitational anomaly.  In the following, this
four-dimensional barotropic hydrodynamic theory will be referred to as
the ``single-fluid'' theory. Related ideas and constructions have also
appeared in other works  \cite{Son:2009tf,Jensen-transgr,Dubovsky:2011sk,
  Haehl:2013hoa,haehl2015adiabatic,galitski,vozmediano-rev}.

Two additional fluid actions were previously introduced on the basis
of the bulk-boundary correspondence associated with anomaly inflow
\cite{abanov2024hydrodynamics}. They were proposed, respectively, as
hydrodynamic descriptions of a single Weyl fermion, referred to as
``Weyl hydrodynamics'', and of a Dirac fermion with independent vector
and axial fluid variables, referred to as ``two-fluid
hydrodynamics''. These theories were not further analyzed, because,
among other things, their actions involve four- and five-dimensional
terms.

In the present work, we confirm and extend these results along the
following lines.
\begin{itemize}
\item We provide a path-integral argument showing how fluid actions
  emerge from the Dirac fermion theory.  After analyzing the anomalies
  in the presence of vector and axial background fields, we rewrite
  the path integral in terms of currents as dynamical fields,
  using a Hubbard--Stratonovich transformation. We
  then assume that the theory enters, in the infrared, into a fluid phase
  with residual current-current interactions. Within the semiclassical
  approximation, we obtain a bosonic low-energy effective action.
\item
  In the simplest case of one vector current, we reobtain
  the hydrodynamic ``single-fluid'' action of our previous work
  \cite{abanov2024hydrodynamics}.
Its dynamical vector field is the fluid momentum,
Legendre transform of the current, while the needed pseudo-scalar field
is now shown to follow from the Abelian Wess--Zumino action
\cite{treiman2014current}.
\item
  The same path-integral construction with two, vector and axial,
  current fields leads to the other actions introduced earlier
  \cite{abanov2024hydrodynamics}.  Gauge symmetry uniquely determines
  their anomalous terms as being the transgression forms
  \cite{mora2006transgression}. These are generalized Chern--Simons
  forms for a pair of gauge fields transforming in the same way,
  identified here as the dynamical fluid momentum and the
  corresponding background field. As a result, the actions naturally
  take the form of combined four- and five-dimensional functionals.
\item
  The correspondence with transgression forms also reveals the
  possibility to extend the single-fluid and two-fluid actions by
  terms which were absent in our earlier work. These are
  gauge-invariant four-dimensional polynomials, which involve two free
  dimensionless couplings.  Although qualitative features, anomalies
  in particular, do not depend on these terms, they may still affect
  physical observables such as transport coefficients.
\item
  The effective field-theory actions obtained from the path
  integral do not by themselves define the corresponding
  hydrodynamics, unless the associated variational principle is also
  specified. In the final part of this work, we address this
  issue. Euler hydrodynamics is known to be a constrained system
  \cite{schutz1977variational,jackiw-rev}, and accordingly the
  fluid action is extremized
  only with respect to restricted, ``admissible'' variations, namely
  diffeomorphisms and gauge transformations of the fluid variables,
  which reflect its basic symmetries
  \cite{schutz1977variational,arnold2008topological,AbanovII}.
  It turns out that the same
  class of variations is required in the present setting. Technically,
  they correctly provide local four-dimensional equations of motion
  from the five-dimensional Chern--Simons terms. Physically, they
  correspond to the fluid motions that remain after hydrodynamic
  relaxation and should therefore be imposed in the effective
  field-theory description.
\end{itemize}

The paper is organized as follows. In Section~2, we obtain the
single-fluid action from the Dirac path integral, after introducing
the Wess--Zumino term and integrating out the fermions in the
semiclassical approximation. In Section~3, we show that the anomalous
part of the action has the form of a transgression, as a consequence of
gauge symmetry. In Section~4, we extend the same path-integral
construction to the Dirac two-fluid theory and also obtain Weyl
hydrodynamics by chiral decomposition. In Section~5, we discuss the
variational principle for the fluid actions and the relation between
hydrodynamic and field-theoretic low-energy theories. Finally, in the
Conclusions, we outline possible developments, such as the
inclusion of temperature and entropy and the study of transport
properties in the hydrodynamic theories considered here.

\section{Hydrodynamics from the fermionic path integral} 
 \label{sec:hydropath}

\subsection{Chiral anomalies and the Wess-Zumino-Witten action}

We begin by recalling the basic facts about chiral anomalies in four dimensions that will be used in the following, adopting the conventions of Ref.~\cite{abanov2024hydrodynamics}. The Dirac fermion in the presence of vector and axial background gauge fields $A_\mu$ and $\tilde A_\mu$ is described by the partition function
\begin{align}
    Z[A,\tilde A]
    =
    \int \bigl[{\cal D}\ppsi\,{\cal D}\bar\ppsi\bigr]_{\slashed D_A}  \exp\!\left(iS_F[\ppsi,A,\tilde A]\right)
    =
    \exp\!\left[\Tr^{\slashed D_A}\ln\!\left(i\slashed D_A\right)\right].
 \label{Z-100}
\end{align}
The Dirac operator and the corresponding action are given by
\begin{align}
    S_F[\ppsi,A,\tilde A]
    &=
    \int_{{\cal M}_4}\bar\ppsi\, i\slashed D_A\,\ppsi
    =
    \int_{{\cal M}_4}\bar\ppsi\,\gamma^\mu
    \left(i\partial_\mu + A_\mu + \gamma^5\tilde A_\mu\right)\ppsi \,.
 \label{L-100}
\end{align}
The notation for the functional measure, $\bigl[{\cal D}\ppsi\,{\cal D}\bar\ppsi\bigr]_{\slashed D_A}$, and for the trace, $\Tr^{\slashed D_A}$, emphasizes that the heat-kernel regularization is defined with respect to the spectrum of the operator $\slashed D_A$.\footnote{For simplicity, the dependence on $\tilde A$ is left implicit in this notation.} This point will become relevant later, when the fermionic determinant is rewritten in terms of hydrodynamic variables while keeping the same regularization scheme.

As is well known, the chiral anomaly reflects the impossibility of
preserving both vector and axial symmetries at the quantum level.
Choosing the heat-kernel regularization so as to preserve vector gauge
symmetry, one finds that the partition function transforms as\footnote{
In the following, we shall often use differential-form notation, for
example $A=A_\mu dx^\mu$.}
\be
\label{Z-trans-100}
Z[\tilde A+d\tilde\lambda,A+d\lambda]
=
Z[\tilde A,A]\,
\exp\!\left(
i\int_{{\cal M}_4}
\tilde\lambda\left(\gamma\, dA\,dA+\alpha\, d\tilde A\,d\tilde A\right)
\right).
\ee
This is the four-dimensional axial anomaly, which contains two
independent terms allowed by time-reversal symmetry. For a single
Dirac fermion, the corresponding coefficients are $\gamma=1$ and
$\alpha=1/3$, in the convention $e/2\pi\to 1$ used in
Ref.~\cite{abanov2024hydrodynamics}. In the following, we set
$\gamma=1$ and keep $\alpha$ explicit for later convenience.

The transformation \eqref{Z-trans-100} implies the Ward identities for
the vector and axial consistent currents,\footnote{In the presence of
both vector and axial backgrounds, there are two kinds of currents,
consistent and covariant, whose different physical meaning is
discussed e.g. in \cite{arouca2022quantum} (See App. D).}
\begin{align}
  j^\mu &= \bar\ppsi \gamma^\mu \ppsi \,,
  \qquad\qquad\ \ \partial_\mu j^\mu = 0 \,,
 \label{jL-100}
 \\
  \tilde j^\mu &= \bar\ppsi \gamma^\mu \gamma^5 \ppsi \,,
  \qquad\qquad \partial_\mu \tilde j^\mu
  = -[dA\,dA + \alpha\, d\tilde A\,d\tilde A] \,.
 \label{jtL-100}
\end{align}
These take the expected form.\footnote{This matches
$j\to J_{\mathrm{cons}}$ and $\tilde j\to \tilde J_{\mathrm{cons}}$ in
Ref.~\cite{abanov2024hydrodynamics}.} Here the square brackets denote
the Hodge dual of a four-form; for example,
\[
[dA\,dA]
=
\epsilon^{\mu\nu\lambda\rho}\,
\partial_\mu A_\nu\,\partial_\lambda A_\rho \,.
\]

It is customary to restore gauge invariance of an anomalous
four-dimensional theory by introducing a five-dimensional topological
``bulk theory,'' so that the combined \(4d\)–\(5d\) system becomes gauge
invariant. This is the anomaly-inflow paradigm, which is physically
realized in topological phases of matter \cite{arouca2022quantum}.
In the present
discussion, no additional physical assumption is needed in order to
add a five-dimensional Chern--Simons term to the partition function
\eqref{Z-100}, namely
\be
 \label{Z-CS-100}
    Z'[A,\tilde A]=Z[A,\tilde A] \exp\left(- i\int_{{\cal M}_5}
    \tilde A dAdA +\a \tilde A d\tilde A d\tilde A \right).
\ee
Here the original four-dimensional spacetime \({\cal M}_4\) is the
boundary of the five-dimensional manifold \({\cal M}_5\),
\(\partial{\cal M}_5={\cal M}_4\). The background fields are assumed
to extend into the extra dimension, but this introduces no ambiguity,
since the gauge variation of the Chern--Simons action is a boundary
term and is therefore independent of the particular extension.

The new partition function \(Z'\) is fully gauge invariant. We may
therefore enlarge the theory by introducing a new pseudo-scalar field
\(\psi(x)\), which realizes axial gauge transformations, and then
integrate over its orbit. This leads to
\begin{align}
  Z''[A,\tilde A]
  &=
  \int {\cal D}\psi\, Z'[A,\tilde A-d\psi]
  \nonumber\\
  &=
  \int {\cal D}\psi\,
  Z[A,\tilde A-d\psi]\,
  \exp\left[
    i\int_{{\cal M}_5}
    (d\psi-\tilde A)\Big(dA\,dA+\alpha\, d\tilde A\,d\tilde A\Big)
  \right].
  \label{Zpp-100}
\end{align}

The new degree of freedom \(\psi\) may at first seem redundant, but it
allows one to introduce the Wess--Zumino--Witten (WZW) action \cite{treiman2014current}. We now
impose the condition
\be
 \label{gc-100}
    \tilde A-d\psi
    \qquad\qquad
    {\rm gauge\ invariant},
\ee
so that \(\psi\) plays the role of a compensating field. In
particular, the partition function
\(Z[A,\tilde A-d\psi]\) in \eqref{Z-100} becomes gauge invariant under
\be
 \label{Apsi-trans-100}
 \tilde A\ \to\ \tilde A+d\tilde\lambda,
 \qquad
 \psi\ \to\ \psi+\tilde\lambda \;.
\ee
The resulting expression represents the gauge-invariant part of the
effective potential obtained from the fermionic determinant,
\(\Gamma=\log Z\). In general, this is a nonlocal and 
unknown functional. We shall expand it in the low-energy limit,
 to leading order in the fields
 and their derivatives, assuming a small mass scale.
 For now let us  take \(\Gamma\simeq 0\),
 i.e. \(Z[A,\tilde A-d\psi]\simeq 1\), and discuss additional terms  later.

In conclusion, we obtain the low-energy effective theory
\(S_{\rm eff}\) for the anomalous Dirac fermion,
\begin{align}
 \label{WZW-100}
    Z'' &= \int {\cal D}\psi\, \exp\Big( iS_{\rm eff}[\psi]\Big),
    \nonumber\\
    S_{\rm eff}[\psi]
    &= \int_{{\cal M}_4}
    \psi\Big( dA\,dA +\alpha\, d\tilde A\, d\tilde A \Big)
    - \int_{{\cal M}_5}
    \tilde A\, dA\, dA+\alpha\, \tilde A\, d\tilde A\, d\tilde A .
\end{align}
This is the gauged Abelian Wess--Zumino--Witten action together with
the bulk background term. The full action is invariant under the gauge
transformations \eqref{Apsi-trans-100} and \(A\to A+d\lambda\).
Moreover, \eqref{WZW-100} correctly reproduces the anomalous
conservation law \eqref{jtL-100} of the four-dimensional boundary
current upon variation with respect to \(\psi\).

It should be noted that the Wess--Zumino--Witten action
\eqref{WZW-100} is a well-known result, which can be obtained in
several ways \cite{preskill1991gauge,treiman2014current}.
It could therefore have been introduced directly.
Nevertheless, the derivation given above is useful for the discussion
that follows.

\subsection{Derivation of the effective hydrodynamic theory
  at low energy}

Starting from the free Dirac theory \eqref{Z-100} in the ultraviolet,
we now assume a renormalization-group flow driven by a relevant
interaction, which brings the system into a low-energy fluid phase.
This phase is assumed to have a small but nonvanishing gap, sufficient
to justify a hydrodynamic description. Since the gauge backgrounds are
not dynamical, the infrared theory retains the fermionic anomaly
structure of the ultraviolet theory, in accordance with the
't~Hooft anomaly-matching condition.

\begin{figure}[t]
\centering
\begin{tikzpicture}[
  scale=1.5,
  >=stealth,
  flow/.style={
  very thick,
  postaction={
    decorate,
    decoration={
      markings,
      mark=at position 0.55 with {\arrow{Stealth[length=5pt,width=6pt]}}
    }
  }
}
]

  \coordinate (UV) at (0,2.2);
  \coordinate (IR) at (0.55,0);

  \draw[flow]
    (UV)
    .. controls (0.70,1.55) and (0.70,0.65) ..
    (IR);

  \fill (UV) circle (2.4pt);
  \fill (IR) circle (2.4pt);

  \node[above left, xshift=-2pt, yshift=-2pt] at (UV) {$UV$};
  \node[below, xshift=-12pt, yshift=2pt] at (IR) {$IR$};

  \draw[flow]
    (1.2,0.5)
    .. controls (0.95,0.40) and (0.70,0.65) ..
    (IR);
  \node[right, xshift=-8pt, yshift=-10pt] at (1.18,0.93) {$\sigma$};

\end{tikzpicture}
\caption{Schematic renormalization-group flow from the ultraviolet
  theory to an infrared fluid regime. Near the infrared fixed point,
  the residual current-current interaction is shown, with
  coupling~$\sigma$.}
\label{fig:rg-flow}
\end{figure}
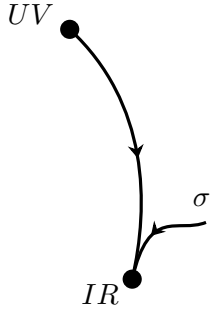

We model the resulting infrared effective theory by supplementing the
fermion theory \eqref{Z-100} with a residual irrelevant
current-current interaction,\footnote{This is in the same spirit as
the Fermi theory of weak interactions. The explicit dependence  $\varepsilon(j)$ is not important for our purposes.}
\be
\label{Sint-100}
S_{int}
=
-\int_{{\cal M}_4}\eps(j)
=
-\int_{{\cal M}_4}\sigma\, j^\mu j_\mu \,,
\qquad\qquad
j^\mu=\bar \ppsi \gamma^\mu \ppsi\,,
\ee
where \(\sigma=O(M^{-2})\) in terms of the infrared mass scale \(M\).
The RG flow just described is shown schematically in
Fig.~\ref{fig:rg-flow}.

The theory defined by the action \(S_F+S_{int}\), with
\(S_F\) and \(S_{int}\) given in \eqref{L-100} and \eqref{Sint-100},
can be reformulated in terms of currents suitable for the
hydrodynamic description, and fermions can be integrated out again.
To this end,
we impose the relation \(j^\mu=\bar\ppsi\gamma^\mu\ppsi\) with the help
of a Lagrange multiplier \(p_\mu\), and write
\begin{align}
    Z[A,\tilde A]
    &=
    \int [{\cal D}\ppsi\,{\cal D}\bar\ppsi]_{\slashed{D}_A}\,
    e^{i\int \bar\ppsi\, i\slashed{D}_A\, \ppsi-\eps(j)}
    \ {\cal D}p_\mu\,{\cal D}j^\mu\,
    e^{i\int p_\mu(\bar\ppsi \gamma^\mu\ppsi-j^\mu)}
    \nonumber\\
    &=
    \int [{\cal D}\ppsi\,{\cal D}\bar\ppsi]_{\slashed{D}_A}\,
    {\cal D}p\,{\cal D}j\,
    \exp\left[
    i\int
    \bar\ppsi\,\gamma^\mu(i\partial_\mu+p_\mu+A_\mu+\gamma^5\tilde A_\mu)\ppsi
    -\eps(j)-p_\mu j^\mu
    \right].
 \label{leg-100}
\end{align}
Now \(j^\mu\) is treated as an independent field in the path integral.
The fermionic integral is again quadratic, but now involves the new
backgrounds
\be
    \pi_\mu=p_\mu+A_\mu, \qquad\ \tilde A_\mu\,,
\ee
where \(p_\mu\) is a gauge-invariant vector field. We may therefore
repeat the heat-kernel evaluation of the fermionic determinant and
obtain the corresponding Wess--Zumino--Witten term, as in
\eqref{WZW-100}. This leads to the effective action
\be
    S_{eff}[\psi,j,p]
    = \int_{{\cal M}_4}\psi (d\pi d\pi+\a d \tilde A d\tilde A) 
    - \eps(j)-p_\mu j^\mu -\int_{{\cal M}_5} \tilde AdAdA
    + \a \tilde A d\tilde A d\tilde A.
\ee
Note that for \(\eps=0\), integration over \(j\) gives \(p=0\), and
the action reduces to the earlier form \eqref{WZW-100}, thereby
providing a check of \eqref{leg-100}.

In the presence of interactions, i.e. \(\sigma\neq 0\) in
\eqref{Sint-100}, the integration over \(j\) can be carried out at the
semiclassical level. The saddle-point equation for \(j\) is precisely
the Legendre transform from the energy density \(\eps(j)\) to the
pressure $P(p)$,\cite{jackiw-rev}
\[
P(p)=-\eps(j)-p_\mu j^\mu,
\qquad
\frac{\partial P}{\partial p_\mu}=-j^\mu,
\]
which leads to\footnote{Note that classical expression of $\eps(j)$ and then
  $P(p)$ receive corrections from the fermionic determinant, to
  be discussed later. They are gauge invariant. In Appendix~\ref{app:2fluid} we also consider a more general case when $ P$ depends on $d\psi-\tilde A$.  }
\be
S_{eff}[\psi, p]= \int_{{\cal M}_4}\psi (d\pi d \pi +\a d \tilde A
d\tilde A)
    +P(p)-\int_{{\cal M}_5} \tilde AdAdA+ \a \tilde A d\tilde A d\tilde A\;.
 \label{Seff-100}
\ee
We thus arrive at an infrared effective action \(S_{eff}\) written in
terms of the bosonic fields \((\psi,p_\mu)\). However, our derivation
is still somewhat naive and requires a correction. We explain how
this can be done, and further justify it later.
 
Anomaly matching tells us that the effective action \eqref{Seff-100}
should be invariant under the axial gauge transformations
\eqref{Apsi-trans-100}, just as the ultraviolet expression
\eqref{WZW-100} is. However, one finds
\be
\d_{\tilde\lambda} S_{eff}=
\int_{{\cal M}_4} \tilde\lambda \left(d\pi d\pi -d A d A\right).
 \label{var-lt-100}
\ee
To see more precisely what happens, note that \(\pi=p+A\) transforms in
the same way as \(A\) on each local patch, since \(p\) is gauge
invariant:
\be
\label{pi-def-100}
    p=\pi- A \qquad\qquad {\rm gauge\ invariant.}
\ee
This parallels the axial condition \eqref{gc-100}. It follows that the
closed forms \(d\pi\) and \(dA\) belong to the same cohomology class,
and hence their instanton numbers are equal,
\[
\int d\pi\, d\pi = \int dA\, dA .
\]
Therefore \(\delta S_{eff}\) vanishes for constant
\(\tilde\lambda\), but not for a general local gauge parameter
\(\tilde\lambda(x)\).

We now modify the effective action \eqref{Seff-100} by adding
local polynomial terms. This is allowed in
renormalization, and may be used to restore symmetries or other
properties that could have been lost in the regularization of the
functional trace
\(\left[{\cal D}\ppsi\,{\cal D}\bar\ppsi\right]_{\slashed{D}_A}\)
in \eqref{Z-100}. The needed term must be a fourth-order polynomial
in \((\pi,A,\tilde A)\), invariant under vector gauge transformations
when \(\pi\) transforms as \(A\), vanishing for \(\pi=A\), and such as
to cancel the variation \eqref{var-lt-100}. An expression with the required
properties is
\be
\Delta S_{eff}
=
\int_{{\cal M}_4}\tilde A\,(\pi-A)\,d(\pi+A)
\;+\;
{\rm g.i.\ terms},
\label{ct-100}
\ee
up to gauge-invariant terms to be discussed later.

Combining \eqref{Seff-100} and \eqref{ct-100}, we arrive at
the gauge-invariant effective action
\begin{align}
S_{eff}[\pi,\psi]=&
\int_{{\cal M}_4} P(\pi-A) + \tilde A (\pi-A)d(\pi +A)+
                    \psi (d\pi d \pi +\a d \tilde A d\tilde A)
\nl
  & -\int_{{\cal M}_5} \tilde A dA dA+ \a \tilde A d\tilde A d\tilde A\,.
  \label{Seff-200}    
\end{align}
This expression is precisely the anomalous hydrodynamic action
proposed in our earlier work \cite{abanov2024hydrodynamics}, now
derived from the fermionic path integral. This is one of the main
results of this work.

Let us add a few remarks.
\begin{itemize}
\item
  The action \eqref{Seff-200} was previously obtained by extending
  the variational formulation of perfect fluids through anomaly
  inflow. In particular, the additional pseudo-scalar field \(\psi\)
  was introduced on phenomenological grounds. The present derivation clearly
  identifies it with the Abelian Wess--Zumino field.
\item Owing to vector gauge invariance, it is natural to rewrite the
  effective action as \(S_{\rm eff}[\pi,\psi]\), namely to pass from
  the gauge-invariant variable \(p\) to \(\pi=p+A\), which transforms
  in the same way as the background field \(A\)
  (cf.~\eqref{pi-def-100}). In fluid language, \(p\) and \(\pi\) may
  be viewed as the kinetic and canonical momenta, respectively. Note
  also that \(S_{\rm eff}\) receives further contributions from the
  semiclassical low-energy expansion of the fermionic determinant. The
  leading four-dimensional terms are gauge-invariant local polynomials
  built from \(\pi-A\), \(d\pi\), and \(dA\), for example a
  Maxwell-type term for \(\pi\). They may be absorbed into a
  redefinition of the pressure, and their explicit form will not be
  important for the hydrodynamic discussion.
\item
  We have assumed that the anomaly coefficients are fixed by
  't~Hooft anomaly matching. More precisely, the irrelevant
  interaction \eqref{Sint-100} may renormalize the effective action
  \cite{galitski}, but the corrections to anomaly coefficients vanish
  in the infrared limit.
\end{itemize}

In summary, the derivation of the hydrodynamic action from the Dirac
theory relies on the following ingredients:

(i) the Wess--Zumino--Witten action, which expresses the anomaly;

(ii) the renormalization-group invariance of the anomaly;

(iii) the infrared effective theory with a residual irrelevant
current-current interaction.

A technically important step is the addition of a local polynomial term
to restore gauge invariance of the effective action. This point is
discussed in the following analysis.

\paragraph{The two-dimensional case.}

  The derivation in this Section can be repeated in two spacetime dimensions.
  Drawing from the analysis of anomalies \cite{arouca2022quantum},
  and following the same steps, one is lead to the effective action
  \begin{align}
S_{eff}[\pi,\psi]=&
\int_{{\cal M}_2} P(\pi-A) + 2\tilde A (\pi-A)+  2 \psi d\pi
 -2\int_{{\cal M}_3} \tilde A dA\,, 
  \label{Seff-50}    
\end{align}
which is the 2d  analog  of \eqref{Seff-200}. Its form was also
guessed by using anomaly inflow in \cite{abanov2024hydrodynamics}.

In contrast with the four-dimensional case, the current-current
interaction \eqref{Sint-100} is now exactly marginal, leading to
a critical line in the UV \cite{ginspargCFT}.\footnote{The value
  of the   anomaly coefficient varies along the line.}
Thus, there is no RG flow into a fluid phase: actually, the theory is
already in the hydrodynamic regime, as discussed in our previous work
\cite{abanov2024hydrodynamics} (see Sect. 2.2).

As is well known, along the critical line the interacting fermion theory
 can be exactly mapped into a bosonic theory \cite{ginspargCFT}.
 Therefore, hydrodynamics of perfect fluids precisely matches two-dimensional
 bosonization.

 The argument in this Section provides another confirmation of this
correspondence. Specifically, the familiar form of the bosonic action\footnote{
  See Ref. \cite{abanov2024hydrodynamics},  Eq. (2.34), and
  \cite{ginspargCFT}.} is obtained from
\eqref{Seff-50} by solving the $\psi$ equation of motion, 
$d\pi=0$, in terms of the scalar field, $\pi=d\th$, and assuming a
quadratic expression for the pressure
$P(\pi-A)= -(\p_\mu\th -A_\mu)^2$.
The correspondence between effective field theory,
hydrodynamics  and bosonization will be further discussed
in the last part of Section 5.

\subsubsection{Computation of Dirac determinants using weighted traces}
 \label{app:wtraces}

The fermion integration measure in the path integral \eqref{Z-100} was
defined by heat-kernel regularization with respect to the spectrum of
\(i\slashed{D}_A\). This choice does not change when the new
background \(\pi=p+A\) is introduced in \eqref{pi-def-100}. From this
point of view, the derivation of the hydrodynamic action
\eqref{Seff-200} amounts to computing the quantity
\begin{align}
  S_{eff} = -i\Tr_{\slashed{D}_A} \ln (i\slashed{D}_\pi)\, +
  \int_{{\cal M}_4} P(p)
  -\int_{{\cal M}_5} \tilde A dA dA+ \a \tilde A d\tilde A d\tilde A\,.
 \label{wtrace-100}
\end{align}
The first term in this expression is known in the mathematical
literature as a ``weighted trace''
\cite{paycha2001renormalized}.\footnote{We thank M.~Braverman for
bringing this to our attention.}
In the argument given above (see Eq. \eqref{Seff-100}), by contrast,
we used the standard regularization
\(-i\Tr_{\slashed{D}_\pi}\ln(i\slashed{D}_\pi)\) and then corrected it
by adding local polynomial terms (see \eqref{ct-100}).

The difference between weighted traces is discussed, for example, in
Theorem~O of Ref.~\cite{cardona2003tracial}. Applied to the
present case, their result may be written as
\begin{align}
 \label{Wodzicki-100}
  \tr_{\slashed{D}_A}\ln D_\pi = \tr_{\slashed{D}_\pi}\ln D_\pi +
  \frac{1}{2}\, \mbox{res}\, (\ln \slashed{D}_\pi - \ln\slashed{D}_A)^2\,.
\end{align}
The left-hand side contains the weighted trace appearing in
\eqref{wtrace-100}, while the right-hand side consists of the standard
heat-kernel regularization used in \eqref{Seff-100}, together with the
so-called Wodzicki residue. The latter is known to be the integral of
a local polynomial in \(\pi\), \(A\), and their derivatives, although
its explicit form was not given in Ref.~\cite{cardona2003tracial}.

This result provides a qualitative justification for the local polynomial term
added above to restore gauge invariance. More precisely, it suggests
that the difference between the two regularization prescriptions is
indeed local. We do not attempt to compute this term directly by
heat-kernel methods.\footnote{Note that the standard approach
regularizes the first variation of the determinant, rather than the
determinant itself \cite{bertlmann2000anomalies}.}
The following Section presents another argument supporting our derivation.

\section{Hydrodynamic action as a transgression form}
\label{sec:trans}

The hydrodynamic action \eqref{Seff-200} consists of the pressure term
\(P(p)\), which specifies the fluid dynamics through the equation of
state, together with differential-form contributions in four and five
dimensions that encode the coupling to the background fields and the
anomalies. We shall show that the differential-form part is naturally
organized in terms of transgression forms, which are familiar objects
in the study of anomalies 
  \cite{Nakahara:Geometry,mora2006transgression,Hull:1990bm}.
Actually, the action \eqref{Seff-200} may be written as
\be
 \label{Seff-trans-100}
 S_{eff}[\pi,A,\psi,\tilde A]
 =
 \int_{{\cal M}_4} P(\pi-A)
 +
 T_5(\pi,A;d\psi,\tilde A)\,,
\ee
where \(T_5(\pi,A;d\psi,\tilde A)\) denotes the 
transgression for the Abelian anomaly. In the
following, we introduce it and discuss its
properties.\footnote{Transgression
forms have also appeared in hydrodynamics in the context of the
Schwinger--Keldysh doubling of the action \cite{Jensen-transgr,Haehl:2013hoa}.
The two formulations do not appear to be directly related.}

\subsection{Transgression for one background}

The transgression form arises by comparing two connections on the same
bundle \cite{Nakahara:Geometry}. As noted above, the gauge fields \(A\) and \(\pi=p+A\)
transform identically under gauge transformations, so the
corresponding Chern forms differ by an exact form. To make contact
with the four-dimensional anomaly, we consider the third Chern class
in six dimensions, which is the starting point of the standard descent
equations \cite{treiman2014current,bertlmann2000anomalies}. We briefly recall the relevant facts and derive the
transgression form in the case of a single Abelian background,
appropriate to the anomaly of a chiral fermion
\cite{mora2006transgression}. The construction will then be extended
to the Dirac case.

The third Chern form \(F_A^3\) associated with the gauge field \(A\)
is a closed and gauge-invariant six-form,
\be
 \label{dF3-100}
    d F_A^3=0, \qquad \delta_\lambda F_A^3=0,
    \qquad\qquad F_A=dA,
\ee
where the gauge variation is \(\delta_\lambda A=d\lambda\). It
therefore admits a local representative,
\be
 \label{F3-CS5-100}
    F_A^3=d C_5(A),
\ee
which is the Chern--Simons form
\begin{equation}
    C_5(A)=A\,dA\,dA .
\end{equation}
Under a gauge transformation one then has
\be
    0=\delta_\lambda F_A^3
    =\delta_\lambda d C_5(A)
    =d\bigl(\delta_\lambda C_5(A)\bigr).
\ee
Thus \(C_5\) is not itself gauge invariant; rather, it transforms by an
exact form,
\begin{equation}
    \delta_\lambda C_5(A)=d\,w_4(A,\lambda),
\end{equation}
where \(w_4(A,\lambda)\) is only locally defined.

We now consider the difference of Chern forms for the
two backgrounds \(\pi\) and \(A\), which transform in the same way
under gauge transformations, as in \eqref{pi-def-100}.
This difference  defines the transgression form \(T_5\),
\be
 \label{T5-def-100}
 F^3_\pi-F^3_A=d T_5(\pi,A)\,.
\ee
Since \(dA\) and \(d\pi\) differ by an exact form, \(T_5\) is globally
defined. It can be obtained by the standard interpolation method
\cite{treiman2014current}, reviewed in Appendix~\ref{app:trans}. The result is
\begin{align}
    T_5(\pi,A)
    &= \pi d\pi d\pi -A dA dA +d\left[A(\pi-A) d(\pi+A)\right]
 \nl
    &= C_5(\pi)-C_5(A) +dB_4(\pi,A)\,.
 \label{T5-200}
\end{align}
Thus the transgression contains the term \(dB_4(\pi,A)\), which
cannot be expressed as the difference of functions of a single
background. Here one already sees the characteristic combination of
four- and five-dimensional terms that appears in the hydrodynamic
action \eqref{Seff-200}.

Note that  when one background is set to zero,
for example \(A=0\), then \(T_5(\pi,0)=C_5(\pi)\), and gauge
invariance is lost. 
Note also that the definitions of \(C_5\) and \(T_5\), given in
Eqs.~\eqref{F3-CS5-100} and \eqref{T5-def-100}, respectively,
determine them only up to exact forms.  These ambiguities are harmless
in the present physical setting. In Appendix~\ref{app:trans}, we discuss this point
in more detail and show that the result \eqref{T5-200} is unique.

\subsection{Transgression for vector and axial backgrounds}

In the hydrodynamic effective action \eqref{Seff-200}, there are two
background fields and two independent parity-invariant anomaly terms,
corresponding to the Chern--Simons forms
\(\tilde A\, d\tilde A\, d\tilde A\) and \(\tilde A\, dA\, dA\).
The corresponding transgression forms involve two pairs of gauge
fields,
\be
 \label{AAt-100}
 (\pi,A), \quad (\tilde \pi,\tilde A), \qquad {\rm with} \qquad \pi-A
 \ \ \ {\rm and}\ \ \ \tilde \pi -\tilde A \qquad {\rm gauge\ invariant}.
\ee
Accordingly, we must enlarge the set of variables considered so far by
promoting the Wess--Zumino pseudoscalar field \(\psi\) to an axial one-form \(\tilde\pi\), in agreement with the gauge-invariance
condition \eqref{Apsi-trans-100},
\be
 \label{psitotilde}
    d\psi \ \ \to\ \ \tilde\pi.
\ee

For the anomaly term \(\tilde A\, d\tilde A\, d\tilde A\), the corresponding
transgression is obtained directly from the single-background formula
\eqref{T5-200}.
This gives
\begin{align}
    T_5^{AAA}(\tilde\pi,\tilde A)
    =
    \tilde \pi d\tilde\pi d\tilde\pi
    -\tilde A d\tilde A d\tilde A
    +
    d\Big[\tilde A(\tilde \pi-\tilde A)d(\tilde\pi+\tilde A)\Big]\,.
 \label{T5AAA-100}
\end{align}

For the anomaly term \(\tilde A\, dA\, dA\), one starts from the mixed
six-dimensional Chern form \(F_{\tilde A}F_A^2\) and integrates its
variation, as described in Appendix~\ref{app:trans}. The result is
\begin{align}
    T_5^{AVV}(\pi,A;\tilde\pi,\tilde A)
    ={}& \tilde \pi d\pi d\pi -\tilde A dA dA
    + d\Big[\tilde A(\pi-A)d(\pi+A)\Big] .
  \label{T5AVV-100}
\end{align}
This expression is manifestly gauge invariant under both vector and
axial gauge transformations \eqref{AAt-100}. However, it is not unique.
In Appendix~\ref{app:trans}, we show that the following four-dimensional
gauge-invariant term can be added, 
\begin{align}
  \label{Q-pol}
T^{AVV}_5 \to T^{AVV}_5 + d Q_{\b,\g},\qquad \quad   Q_{\beta,\gamma}
    =  (\tilde \pi-\tilde A)(\pi-A)
    d(\beta(\pi-A) +\gamma A)\,,
\end{align}
parameterized by the two dimensionless constants
 \((\beta,\gamma)\). This is the
freedom of defining the transgression up to exact gauge-invariant terms.
A physical interpretation of this fact will
emerge later from the decomposition in chiral components
\(A_\pm=A\pm\tilde A\) and \(\pi_\pm=\pi\pm\tilde\pi\); see
Section~\ref{sec:2fEoM}.

\subsection{Rewriting the hydrodynamic single-fluid action}

We now have the ingredients needed to rewrite the single-fluid
hydrodynamic action in terms of transgression forms, as anticipated in
\eqref{Seff-trans-100}. Using the expressions
\eqref{T5AAA-100}, \eqref{T5AVV-100} an \eqref{Q-pol}, evaluated at
\(\tilde\pi=d\psi\), we obtain
\begin{align}
  S_I
  &=
  \int_{{\cal M}_4}P(\pi-A)+
  \int_{{\cal M}_5}T_5^{AVV}(\pi,A;d\psi,\tilde A)
  +\alpha\,T_5^{AAA}(d\psi,\tilde A) +d Q_{\beta, \gamma}(\pi,A;d\psi,\tilde A)
  \nonumber\\
  &=
  \int_{{\cal M}_4}P(\pi-A)
  +\tilde A(\pi-A)d(\pi+A)
  +\psi(d\pi d\pi+\alpha\, d\tilde A d\tilde A)
  \nonumber\\
  &\quad
  +\int_{{\cal M}_4}(d\psi-\tilde A)(\pi-A)d(\beta(\pi-A)+\gamma A)
  -\int_{{\cal M}_5}\tilde A dA dA+\alpha\,\tilde A d\tilde A d\tilde A\,.
  \label{Seff-psi-200}
\end{align}
All terms in the action \eqref{Seff-200} are thus recovered, together
with the gauge-invariant polynomial left undetermined in
\eqref{ct-100}. Note that, upon setting \(\tilde\pi=d\psi\), all
dynamical terms involving \((\pi,\tilde\pi)\) become
four-dimensional; only the pure background terms remain
five-dimensional.

Thus, we have shown how the action \eqref{Seff-psi-200} entirely arises from
the anomalous gauge structure of the Dirac path integral. It has the
same form as the hydrodynamic action proposed earlier
\cite{abanov2024hydrodynamics}, supplemented by the gauge-invariant
term $Q_{\b,\g}$ with two free coupling constants.

\section{ Actions for two-fluid and Weyl fluid hydrodynamics}

\subsection{Two-fluid hydrodynamics from the path integral}

The discussion of transgression forms in the previous Section shows
that the anomalous action naturally involves two pairs
of fields, \((\pi,A)\) and \((\tilde\pi,\tilde A)\), one pair for each
background and gauge symmetry. This suggests a generalized
hydrodynamic theory with two independent components, vector and axial,
described by the corresponding momenta \(\pi\) and
\(\tilde\pi\).

The following action is found by analogy with \eqref{Seff-psi-200},
\begin{align}
  S_{II}
  =&
  \int_{{\cal M}_4} P(\pi-A,\tilde \pi - \tilde A) + Q_{\beta,\gamma}(\pi,A;\tilde \pi,\tilde A)
  + \int_{{\cal M}_5} T_5^{AVV}(\pi,A;\tilde\pi,\tilde A)
  + \alpha\, T_5^{AAA}(\tilde\pi,\tilde A)
\nonumber  \\
  =&
  \int_{{\cal M}_4} P(\pi-A,\tilde \pi - \tilde A)
  + \tilde A(\pi-A)d(\pi+A)
  + \alpha\,\tilde A(\tilde \pi-\tilde A)d(\tilde\pi+\tilde A)
  \nl
  &
  + \int_{{\cal M}_4} (\tilde \pi-\tilde A)(\pi-A)
    d(\beta(\pi-A) +\gamma A)
  \nonumber\\
  & + \int_{{\cal M}_5}\tilde \pi d\pi d\pi -\tilde A dA dA
  +  \int_{{\cal M}_5} \alpha\,\tilde \pi d\tilde\pi d\tilde\pi
  - \alpha\, \tilde A d\tilde A d\tilde A \,,
  \label{Seff-II-100} 
\end{align}
which reduces to \eqref{Seff-psi-200} upon setting
\(\tilde\pi=d\psi\).

A new feature  is that the
dynamical fields \(\pi\) and \(\tilde\pi\) now extend into the
five-dimensional bulk, through the Chern--Simons terms in the last
line of \eqref{Seff-II-100}.\footnote{As noted earlier, extending the
background fields \(A\) and \(\tilde A\) into the fifth dimension
poses no difficulty.} As the previous discussion makes clear, these
topological terms are essential for a proper treatment of the
anomalies.

In the following, we show that the two-fluid action \(S_{II}\) in
\eqref{Seff-II-100} can likewise be obtained from the fermionic path
integral in the presence of an irrelevant infrared interaction, by
extending the argument of Section~\ref{sec:hydropath}.

We again start from the free fermion theory \eqref{Z-100}, made gauge
invariant by the addition of the Chern--Simons background terms, as in
\(Z'[A,\tilde A]\) in \eqref{Z-CS-100}. We then introduce independent
vector and axial currents, which can describe more general
current-current interactions, as follows
\be
 \label{epsjtj-100}
 j^\mu=\bar\ppsi \gamma^\mu \ppsi, \qquad\quad
 \tilde j^\mu=\bar\ppsi \gamma^\mu \gamma^5 \ppsi,\qquad\quad
 \eps=\eps(j,\tilde j).
\ee
The current variables are introduced into the path integral with
the help of gauge-invariant Lagrange multipliers \(p\) and \(\tilde p\),
in direct analogy with \eqref{leg-100},
\begin{align}
 \label{Zp-leg-200}
  \int [{\cal D}\ppsi\,{\cal D}\bar\ppsi]_{\slashed{D}_A} 
    {\cal D}p\,{\cal D}j\,{\cal D}\tilde p\,{\cal D}\tilde j\,
    \exp\left[
    i\int \bar\ppsi \gamma^\mu(i\partial_\mu +p_\mu+A_\mu 
    + \gamma^5(\tilde A_\mu+\tilde p_\mu))\ppsi
    -\eps(j,\tilde j)
    - p_\mu j^\mu -\tilde p_\mu \tilde j^\mu
    \right].
 \nonumber
\end{align}
Semiclassical integration over \(j\) and \(\tilde j\) then amounts to a
double Legendre transform,
\be
 \label{pressure-200}
    P(p,\tilde p)=-\eps(j,\tilde j)- p_\mu j^\mu -\tilde p_\mu \tilde j^\mu, 
    \qquad \quad \frac{\d P}{\d p_\mu} =-j^\mu,
    \qquad \frac{\d P}{\d \tilde p_\mu} =-\tilde j^\mu \,.
\ee
Thus the pressure becomes a function of the two 
variables \(p\) and \(\tilde p\).

The fermionic determinant is now evaluated in the presence of the new
backgrounds \(\pi=p+A\) and \(\tilde\pi=\tilde p+\tilde A\).
A naive repetition of the previous argument would then lead to the
effective action
\begin{align}
  Z'[A,\tilde A]
  &= \exp\bigl(i S_{eff}[\pi,A,\tilde\pi,\tilde A]\bigr),
  \\
  S_{eff}[\pi,A,\tilde\pi,\tilde A]
  &= \int_{{\cal M}_4} P(p,\tilde p)
  + \int_{{\cal M}_5}
  \tilde \pi\, d\pi\, d\pi
  + \a\,\tilde \pi\, d\tilde\pi\, d\tilde\pi
  - \tilde A\, dA\, dA
  - \a\, \tilde A\, d\tilde A\, d\tilde A \,,
\end{align}
where the anomaly is represented in terms of Chern--Simons forms
rather than the Wess--Zumino action \eqref{WZW-100}, since the full
axial field \(\tilde\pi\) is now available.

As in the single-fluid case (cf.~\eqref{Seff-100}), this expression is
too naive: it is only globally gauge invariant, not locally so.
As before, one must therefore allow for additional polynomial terms
that restore local gauge invariance. In light of the previous
analysis, we now recognize that the appropriate gauge-invariant object
built from Chern--Simons terms associated with fields in the same
cohomology class, namely \((\pi,A)\) and \((\tilde\pi,\tilde A)\), is
the transgression form. In this way, we confirm the previously
conjectured expression \eqref{Seff-II-100}.

As discussed in Section 2.2.1, we view
\(S_{II}\) in \eqref{Seff-II-100} as the leading low-energy part of
the ratio of fermion determinants, evaluated before and after the
inclusion of the irrelevant interaction \eqref{epsjtj-100}, with the
regularization prescription kept fixed.

This completes the derivation of the two-fluid hydrodynamic action
\eqref{Seff-II-100} from the fermionic path integral. This action
was guessed in \cite{abanov2024hydrodynamics},
although the gauge-invariant term \(Q_{\beta,\gamma}\)
was not included there.

\subsection{Weyl-fluid action}

The hydrodynamic theory of a single Weyl fermion follows naturally from
the two-fluid theory by chiral decomposition of the gauge fields and
currents, corresponding to the splitting of a Dirac fermion into
Weyl components. We introduce
\be
 \label{Aplusminus}
 A_\pm=A\pm \tilde A,
 \qquad
 j_\pm=\frac{1}{2}(j \pm \tilde j),
 \qquad\to\qquad
 j^\mu A_\mu + \tilde j^\mu \tilde A_\mu
 =
 j^\mu_+ A_{+\mu} +j^\mu_- A_{-\mu}\,,
\ee
and similarly for \(\pi_\pm=\pi \pm \tilde\pi\).

The transgression forms for the vector and axial sectors,
\eqref{T5AAA-100} and \eqref{T5AVV-100}, also admit a chiral
decomposition,
\begin{align}
    T_5^{AVV}(\pi,A;\tilde\pi,\tilde A)
    + \frac{1}{3}\,T_5^{AAA}(\tilde\pi,\tilde A)
    + dQ_{2/3,1}(\pi,A;\tilde\pi,\tilde A)
    =
    \frac{1}{6}\,T_5(\pi_+,A_+)
    - \frac{1}{6}\,T_5(\pi_-,A_-) \,,
 \label{chir-split-100}
\end{align}
where $T_5(\pi_\pm,A_\pm)$ have the form \eqref{T5AAA-100}.
The identity \eqref{chir-split-100} requires some algebra to verify,
but it is natural in view of the chiral decomposition of the
Chern--Simons terms into \(C_5(A_\pm)\).
For the Dirac fermion, the anomaly coefficient is
\(\alpha=1/3\). Moreover, the decomposition
\eqref{chir-split-100} holds for the choice
\((\beta,\gamma)=(2/3,1)\) of the gauge-invariant term
\(Q_{\beta,\gamma}\) in \eqref{Seff-II-100}. For more general values
of these parameters, there is an additional term 
coupling the two chiralities, to be discussed later in
Section~\ref{sec:2fEoM}.

The Weyl-fluid action then takes the form
\begin{align}
  \label{S-weyl}
    S_W
    &=
    \int_{{\cal M}_4} P(\pi-A) + \frac{1}{6}T_5(\pi,A)
 \\
    &=
    \int_{{\cal M}_4} P(\pi-A)
    + \frac{1}{6}\,A(\pi-A)d(\pi+A)
    + \frac{1}{6}\int_{{\cal M}_5} \pi\, d\pi\, d\pi - A\, dA\, dA \,,
 \nonumber
\end{align}
after identifying, e.g. \((\pi_+,A_+)\to(\pi,A)\). This action
can likewise be obtained from the fermionic path integral by following
the same steps as in the previous subsection. It does not involve free
parameters, since the coefficient $1/6$ is fixed by anomaly.

We again remark that the Weyl-fluid theory involves dynamical
five-dimensional terms, and therefore cannot be formulated as a local
four-dimensional theory by itself. The corresponding variational
principle, and the way in which it leads to local four-dimensional
hydrodynamic equations, is discussed in the next Section.

\section{Variational principle for fluid actions}
\label{sec:eom}

The derivation of the hydrodynamic actions from the path integral is
not complete without specifying the corresponding variational
principle. In Euler (ideal) hydrodynamics, the action is not varied freely
over all configurations of the fluid momentum \(\pi\), since the
underlying Hamiltonian system is constrained \cite{jackiw-rev}.
The ``admissible''
variations are instead the restricted ones generated by gauge
transformations and diffeomorphisms, which correspond to the basic
symmetries of fluids \cite{schutz1977variational,arnold2008topological}. 
In standard variational
formulations of Euler hydrodynamics, these restrictions are often
encoded by introducing Clebsch variables or similar auxiliary fields
\cite{jackiw-rev,mobbs1982variational}.
In the present setting, it is more convenient to keep the momentum
one-form \(\pi\) as the basic variable and impose the admissible
variations directly.

This restricted variational principle is physically motivated
\cite{schutz1977variational}.
At the end of this Section, we shall argue that this is a necessary
ingredient in going from effective field theory to hydrodynamics, and that
it amounts to a projection into a subspace of motions allowed after
hydrodynamic equilibration.

In the following, we shall apply the variations of
gauge transformations and diffeomorphisms 
to the Weyl, two-fluid and single-fluid actions.
We shall find consistent results in the presence of anomalous
terms, which actually respect these symmetries.
Furthermore, restricted variations will allow us to obtain
local four-dimensional equations of motion from these actions which
contain five-dimensional Chern--Simons functionals.

\subsection{Equations of motion for the Weyl action}

We begin with the Weyl fluid action, which already contains the main
features of the mixed 4d-5d variational problem. For
convenience, we repeat it here,\footnote{The factor \(1/6\) is omitted
by rescaling \(6P\to P\); it can be restored in the final expressions.}
\begin{align}
  \label{S-weyl2}
    S_W = S_4+S_5
    = \int_{{\cal M}_4} P(\pi-A) + A(\pi-A)d(\pi+A)
      + \int_{{\cal M}_5} \pi\,d\pi\,d\pi - A\,dA\,dA \,,
\end{align}
where \({\cal M}_4=\partial{\cal M}_5\). The action has been written
as the sum of its four- and five-dimensional parts in order to make
the boundary contributions to the variation explicit.

\paragraph{General variation.}
We begin by varying the Weyl action with respect to both \(\pi\) and
\(A\), treating them as fields defined on the full 4d-5d
space. The variation of the five-dimensional Chern--Simons terms
produces both bulk and boundary contributions. For example,
\be
\label{b-t}
\delta\bigl(\pi\,d\pi\,d\pi\bigr)
=
3\,\delta\pi\,d\pi\,d\pi
+
2\,d\bigl(\delta\pi\,\pi\,d\pi\bigr)\,,
\ee
and similarly for the term \(A\,dA\,dA\). The boundary contribution
must then be combined with the variation of the four-dimensional part
of the action. In this way one obtains
\begin{align}
  \delta S_W
  &=
  \int_{{\cal M}_4}
  \delta A_\mu\, J^\mu_{\rm cons}
  -
  \delta \pi_\mu\, {\cal J}^\mu_{\rm cons}
  +\delta\!\left(\int_{{\cal M}_5}\pi\,d\pi\,d\pi-A\,dA\,dA\right)
 \nonumber\\
  &=
  \int_{{\cal M}_4}
  \delta A_\mu\, J^\mu_{\rm cov}
  -
  \delta \pi_\mu\, {\cal J}^\mu_{\rm cov}
  +3\int_{{\cal M}_5}\delta\pi\,d\pi\,d\pi
  -3\int_{{\cal M}_5}\delta A\,dA\,dA,
 \label{S-weyl-var}
\end{align}
where
\begin{align}
  J^\mu_{\rm cov}
  &=
  n^\mu +[(\pi-A)d(\pi+2A)]^\mu ,
  \qquad\quad
  n^\mu=-\frac{\partial P}{\partial p_\mu} \,,
  \label{Jcov-500}\\
  {\cal J}^\mu_{\rm cov}
  &=
  n^\mu -[(\pi-A)d(2\pi+A)]^\mu ,
 \label{calJcov-500}\\
  J^\mu_{\rm cons}
  &=J^\mu+2[AdA]^\mu,
 \label{J-cons-cov-500}\\
  {\cal J}^\mu_{\rm cons}
  &={\cal J}^\mu+2[\pi d\pi]^\mu .
 \label{calJ-cons-cov-500}
\end{align}
The first line in \eqref{S-weyl-var} corresponds to varying the
four-dimensional part of the action and thus defines the consistent
currents. In the second line, the boundary contribution from the
variation of the five-dimensional term has been included; this yields
the covariant currents, which in the present Abelian case are also
gauge invariant.\footnote{The properties of consistent and covariant
  currents are described in Appendix D of \cite{arouca2022quantum}.}
Here and in what follows, we shall work primarily with 
covariant currents and omit the subscript ``cov'' whenever no confusion
can arise.

To obtain hydrodynamic equations of motion, we restrict the
general variation \eqref{S-weyl-var} to admissible gauge and
diffeomorphism variations.

\paragraph{Gauge variations.}
We begin with the simultaneous gauge transformations of the full
action,
\begin{equation}
\delta_\lambda \pi=\delta_\lambda A=d\lambda\,.
 \label{gvar-500}
\end{equation}
Substituting \eqref{gvar-500} into the general variation
\eqref{S-weyl-var}, one finds
\be
0
=
\int_{{\cal M}_4}\lambda
\left(
\partial_\mu {\cal J}^\mu +3[d\pi\,d\pi]
-\partial_\mu J^\mu -3[dA\,dA]
\right),
\ee
and hence the identity
\be
\label{ident-500}
\partial_\mu J^\mu +3 [dA\,dA]
=
\partial_\mu{\cal J}^\mu +3 [d\pi\,d\pi] \,.
\ee
This relation may also be checked directly from
\eqref{Jcov-500} and \eqref{calJcov-500}.

To obtain the hydrodynamic equation of motion, however, one should
restrict the variation to the dynamical field \(\pi\) while keeping the
background \(A\) fixed. Thus we consider the admissible gauge variation
\begin{equation}
\delta_\lambda \pi=d\lambda,
\qquad\qquad
\delta_\lambda A=0.
\label{gvar-pi-500}
\end{equation}
Using \eqref{gvar-pi-500} in \eqref{S-weyl-var}, we obtain
\be
0
=
\int_{{\cal M}_4}\frac{\delta S_W}{\delta \pi_\mu}\,\delta_\lambda \pi_\mu
=
\int_{{\cal M}_4}\lambda
\left(
\partial_\mu {\cal J}^\mu +3[d\pi\,d\pi]
\right),
\ee
and therefore
\begin{align}
    \partial_\mu {\cal J}^\mu 
    +3 [d\pi\,d\pi] =0.
 \label{dcalJ-500}
\end{align}
This is the gauge-variation equation of motion for the fluid and
expresses conservation of the particle current.

Combining \eqref{dcalJ-500} with the identity \eqref{ident-500}, one
obtains the corresponding equation for the electric current,
\begin{align}
    \partial_\mu J^\mu = -3[dA\,dA].
 \label{dJ-400}
\end{align}
This is precisely the expected anomaly equation. It may equivalently
be obtained by anomaly inflow from the five-dimensional bulk,
\be
\label{inflows}
\partial_\mu J^\mu
=
j^{5}
=
\frac{\delta S_5}{\delta A_5}
=
-3 [dA\,dA]\,,
\ee
where \(j^{5}\) is the five-dimensional current in the direction
normal to the boundary.

\paragraph{Diffeomorphism variations.}
We now turn to the second class of admissible variations, namely
diffeomorphisms acting on the dynamical field \(\pi\),
\be
\delta_\xi \pi_\mu
=
\xi^\alpha \partial_\alpha \pi_\mu
+
\pi_\alpha \partial_\mu \xi^\alpha,
\qquad\qquad
\delta_\xi \pi
=
{\cal L}_\xi \pi
=
\bigl(i_\xi d + d\,i_\xi\bigr)\pi\,.
\label{diffvar-500}
\ee
We restrict to vector fields \(\xi\) tangent to the boundary,
\be
\xi^\perp\big|_{\partial M_5}=0\,,
\label{tangent-bdry-500}
\ee
so that the diffeomorphism preserves \(\partial M_5\).
The variation of the five-dimensional Chern--Simons term is
a boundary term,
\begin{align}
\delta_\xi S_5
&=
\int_{{\cal M}_5}{\cal L}_\xi\bigl(\pi\,d\pi\,d\pi\bigr)
=
\int_{{\cal M}_5}d\,i_\xi\bigl(\pi\,d\pi\,d\pi\bigr)
=
\int_{{\cal M}_4}i_\xi\bigl(\pi\,d\pi\,d\pi\bigr),
\label{diffS5-500}
\end{align}
which vanishes because \(\xi\) has no component normal to the boundary.
Therefore only the four-dimensional part of the action contributes, and
the equation of motion is naturally written in terms of the consistent
current:
\begin{align}
\delta_\xi S_W
&=
-\int_{{\cal M}_4}{\cal J}^\mu_{\rm cons}\,{\cal L}_\xi \pi_\mu
=
\int_{{\cal M}_4}\xi^\alpha
\left(
{\cal J}^\mu_{\rm cons}\,\partial_{[\mu}\pi_{\alpha]}
+
\pi_\alpha\,\partial_\mu{\cal J}^\mu_{\rm cons}
\right).
\label{diffSW-500}
\end{align}
Since \(\xi^\alpha\) is arbitrary on \({\cal M}_4\), stationarity
implies
\begin{align}
{\cal J}^\mu_{\rm cons}\,\partial_{[\mu}\pi_{\alpha]}
+
\pi_\alpha\,\partial_\mu{\cal J}^\mu_{\rm cons}
=0.
\label{CLcons-500}
\end{align}
This is the Carter--Lichnerowicz form of the Euler equation \cite{lichnerowicz1994relativistic,AbanovII}.

Using \eqref{calJ-cons-cov-500}, together with the identity
\[
\pi_\alpha [d\pi\,d\pi]
=
2[\pi\,d\pi]^\mu \partial_{[\mu}\pi_{\alpha]},
\]
one finds that \eqref{CLcons-500} is equivalent to
\begin{align}
{\cal J}^\mu\,\partial_{[\mu}\pi_{\alpha]}
+
\pi_\alpha\bigl(\partial_\mu{\cal J}^\mu +3[d\pi\,d\pi]\bigr)
=0.
\label{CLcov-500}
\end{align}

Using the gauge equation \eqref{dcalJ-500},\footnote{In the barotropic
case, the gauge equation is in fact not independent: contracting
\eqref{CLcov-500} with \({\cal J}^\alpha\) and assuming
\({\cal J}\neq 0\) yields
\(\partial_\mu{\cal J}^\mu +3[d\pi\,d\pi]=0\). This degeneracy is
special to the barotropic theory.} this reduces to
\begin{align}
    {\cal J}^\mu\,\partial_{[\mu}\pi_{\alpha]}=0.
 \label{EulerW-500}
\end{align}
For nonvanishing \({\cal J}\), the equation \eqref{EulerW-500}
implies that the two-form \(d\pi\) does not have maximal rank, and therefore
\begin{align}
d\pi\,d\pi=0,
\label{dpipi-500}
\end{align}
as in the standard four-dimensional Carter--Lichnerowicz
analysis.\footnote{See, for example, Section~3.1.1 of \cite{abanov2024hydrodynamics}.}
Together with the
gauge equation, this implies the conservation of the particle current,
\begin{align}
\label{dcalJ-0}
    \p_\mu {\cal J}^\mu=0.
\end{align}

Finally, it is convenient to rewrite the equations in the standard
force-balance form. Introducing the stress tensor
\begin{align}
T^\mu{}_\nu
=
{\cal J}^\mu p_\nu
+
P\,\delta^\mu{}_\nu,
\qquad\qquad
p_\nu=\pi_\nu-A_\nu,
\label{TW-500}
\end{align}
one finds, by combining the gauge equation \eqref{dcalJ-500} with the
diffeomorphism equation \eqref{EulerW-500}, that
\begin{align}
\partial_\mu T^\mu{}_\nu
=
F_{\nu\mu}J^\mu,
\qquad\qquad
F_{\mu\nu}=\partial_\mu A_\nu-\partial_\nu A_\mu.
\label{TW-force-500}
\end{align}
This is the standard force-balance equation for the Weyl fluid in the
presence of the external gauge field. The current appearing in the
Lorentz-force term is the covariant current \(J^\mu\), as expected.
\footnote{The results in this Section agree with those of
    Ref. \cite{WiegWZ} for $T=0$ and with $\tilde A =0$.}

We have thus shown that the admissible gauge and diffeomorphism
variations can be consistently applied to the mixed 4d-5d Weyl
action \eqref{S-weyl2}. They lead to local four-dimensional equations
of motion: the anomalous conservation law for the electric current and
the Carter--Lichnerowicz equation for the fluid momentum. It is worth
emphasizing that the
reduction from a mixed-dimension action to local 4d equations
does not occur under unrestricted variations. It
is precisely the choice of admissible variations, known in hydrodynamics,
that makes this result possible.

\subsection{Equations of motion for the two-fluid action}
\label{sec:2fEoM}

We now turn to the two-fluid action \eqref{Seff-II-100}. As in the
Weyl case, the action is decomposed into four- and five-dimensional terms,
\begin{align}
  \label{Seff-II-200}
  S_{II}
  =
  S_{II,4}+S_{II,5},
\end{align}
with
\begin{align}
  S_{II,4}
  =&
  \int_{{\cal M}_4} P(\pi-A,\tilde \pi-\tilde A)
  + \tilde A(\pi-A)d(\pi+A)
  + \alpha\,\tilde A(\tilde \pi-\tilde A)d(\tilde\pi+\tilde A)
  + Q_{\beta,\gamma},
  \label{Seff-II-201}
  \\
  S_{II,5}
  =&
  \int_{{\cal M}_5}\tilde \pi\, d\pi\, d\pi
  + \alpha\, \tilde \pi\, d\tilde\pi\, d\tilde\pi
  - \tilde A\, dA\, dA
  - \alpha\, \tilde A\, d\tilde A\, d\tilde A \,,
\label{Seff-II-202}
\end{align}
where \(Q_{\beta,\gamma}\) denotes the gauge-invariant two-parameter
polynomial introduced in \eqref{Q-pol}.\footnote{Explicit expressions
for the currents and related quantities are collected in Appendix~\ref{app:2fluid}. For one Dirac fermion, \(\alpha=1/3\).}

\paragraph{General variation.}
We vary the action with respect to the fluid fields \(\pi\) and
\(\tilde\pi\), treating them as fields on the full \(4d\)–\(5d\)
space. As in the Weyl case, the variation of \(S_{II,5}\) produces
both bulk and boundary terms, and the latter are combined with the
variation of \(S_{II,4}\). This defines the covariant four-dimensional
currents through the full variation (see Appendix~\ref{app:2fluid})
\begin{align}
  \delta S_{II}
  =
  \int_{{\cal M}_4}
  \delta A_\mu\, J^\mu
  + \delta\tilde A_\mu\, \tilde J^\mu
  - \delta\pi_\mu\, {\cal J}^\mu
  - \delta\tilde\pi_\mu\, \tilde{\cal J}^\mu
  \,+\, \delta S_{II,5}^{\rm bulk},
  \label{S-II-var}
\end{align}
where \(\delta S_{II,5}^{\rm bulk}\) only contains the bulk five-dimensional
variations, as in \eqref{S-weyl-var}. In what follows we work with these covariant currents. As
before, one may also define the corresponding consistent currents by
varying \(S_{II,4}\) alone.

\paragraph{Gauge variations.}
We first consider simultaneous gauge transformations of the pairs
\((\pi,A)\) and \((\tilde\pi,\tilde A)\),
\begin{align}
\delta_\lambda \pi=d\lambda,
\qquad
\delta_\lambda A=d\lambda,
\qquad
\delta_{\tilde\lambda}\tilde\pi=d\tilde\lambda,
\qquad
\delta_{\tilde\lambda}\tilde A=d\tilde\lambda.
\label{gvar-II}
\end{align}
Substituting these variations into the full variation gives the following
identities, owing to overall gauge invariance,
\begin{align}
  \partial_\mu J^\mu+2[dA\,d\tilde A]
  &=
  \partial_\mu {\cal J}^\mu+2[d\pi\,d\tilde\pi],
 \nonumber\\
  \partial_\mu\tilde J^\mu+[dA\,dA+3\alpha\,d\tilde A\,d\tilde A]
  &=
  \partial_\mu \tilde{\cal J}^\mu+[d\pi\,d\pi+3\alpha\,d\tilde\pi\,d\tilde\pi].
  \label{J-II}
\end{align}

The hydrodynamic equations of motion are obtained by restricting the
variation to the dynamical fields \(\pi\) and \(\tilde\pi\), while the
backgrounds are kept fixed. Thus we take
\begin{align}
\delta_\lambda \pi=d\lambda,\qquad \delta_\lambda A=0,
\qquad
\delta_{\tilde\lambda}\tilde\pi=d\tilde\lambda,\qquad
\delta_{\tilde\lambda}\tilde A=0.
\label{gvar-II-dyn}
\end{align}
Using \eqref{S-II-var}, we obtain
\begin{align}
    d{\cal J} +2\,d\pi\,d\tilde\pi &= 0,
 \nonumber\\
    d\tilde{\cal J} + d\pi\,d\pi +3\alpha\, d\tilde \pi\,d\tilde \pi &=0.
  \label{J-part-cons}
\end{align}
Combining these equations with the identities \eqref{J-II}, one finds
the anomaly equations for the charge currents,
\begin{align}
 \partial_\mu J^\mu &= -2 [dA\,d\tilde A],
 \nonumber\\
 \partial_\mu\tilde J^\mu &= -[dA\,dA] -3\alpha[d\tilde A\,d\tilde A].
\label{J-II-anom}
\end{align}
As expected, these anomaly equations are independent of the explicit form of
the term \(Q_{\beta,\gamma}\), which only affects the
constitutive expressions for the currents.

\paragraph{Common diffeomorphism variations.}
Let us start by varying the two fluid variables together under a
common diffeomorphism,
\begin{align}
\delta_\xi\pi={\cal L}_\xi\pi,
\qquad
\delta_\xi\tilde\pi={\cal L}_\xi\tilde\pi.
\label{commondiff-II}
\end{align}
Restricting to vector fields tangent to the boundary, as in the Weyl
case, the variation of \(S_{II,5}\) is again a boundary term and
vanishes (cf. \eqref{diffS5-500}). The variation of \(S_{II,4}\) then gives
\begin{align}
  {\cal J}^\mu_{\rm cons}\partial_{[\mu}\pi_{\alpha]}
  +\pi_\alpha\,\partial_\mu {\cal J}^\mu_{\rm cons}
  +\tilde{\cal J}^\mu_{\rm cons}\partial_{[\mu}\tilde\pi_{\alpha]}
  +\tilde\pi_\alpha\,\partial_\mu \tilde{\cal J}^\mu_{\rm cons}
  =0.
\label{common-diff-eq}
\end{align}
Actually, this equation is too weak to determine the two independent
fluid momenta $\pi$ and $\tilde\pi$:  together with the
gauge equations \eqref{J-part-cons}, they make six conditions for
eight variables.\footnote{Note that they do not imply the conservation
  of particle currents \({\cal J}\) and \(\tilde{\cal J}\) as in
  \eqref{dcalJ-0} (no inflow).} We conclude that the common
diffeomorphism variation does not by itself provide a satisfactory
local variational principle for a genuine two-fluid theory.  We shall
further discuss this result at the end of this Section.

\paragraph{Independent diffeomorphisms.}
One may next try to vary \(\pi\) and \(\tilde\pi\) independently, for
example
\begin{align}
\delta_{\tilde\xi}\tilde\pi={\cal L}_{\tilde\xi}\tilde\pi,
\qquad
\delta_{\tilde\xi}\pi=0.
\label{diff-pitilde}
\end{align}
For the term \(\alpha\,\tilde\pi\,d\tilde\pi\,d\tilde\pi\), the
variation is analogous to the Weyl case and reduces to a boundary term.
The mixed term, however, behaves differently:
\begin{align}
  \delta_{\tilde\xi}\int_{M_5} \tilde\pi\,d\pi\,d\pi
  &=
  \int_{M_5} (i_{\tilde\xi}d\tilde\pi)\,d\pi\,d\pi
  +
  \int_{M_5} d\Big[(i_{\tilde\xi}\tilde\pi)\,d\pi\,d\pi\Big]
  \nonumber\\
  &=
  \int_{M_5} (i_{\tilde\xi}d\tilde\pi)\,d\pi\,d\pi
  +
  \int_{M_4} (i_{\tilde\xi}\tilde\pi)\,d\pi\,d\pi .
\label{pitilde-diff-var}
\end{align}
Thus, independent diffeomorphisms of \(\pi\) and \(\tilde\pi\) produce
both bulk and boundary contributions, and do not lead to closed local
equations on \({\cal M}_4\). In this form, the variational problem does not define a local four-dimensional  hydrodynamic theory.

The correct formulation of independent diffeomorphisms is found
by considering the chiral decomposition
\begin{align}
\pi_\pm &=\pi\pm \sqrt{3\alpha}\,\tilde\pi,
\qquad\qquad
A_\pm =A\pm\sqrt{3\alpha}\, \tilde A,
 \label{chir-split}\\
  {\cal J}_\pm &=\frac12\left({\cal J}\pm
                 \frac{\tilde{\cal J}}{\sqrt{3\alpha}}\right),
\quad\;\;
J_\pm =\frac12\left(J\pm \frac{\tilde J}{\sqrt{3\alpha}}\right).
 \nonumber
\end{align}
The five-dimensional action splits
into two chiral Chern--Simons terms,
\begin{align}
S_{II,5}
=
S_+ - S_-,
\qquad
S_\pm
=
C\int_{M_5}\pi_\pm\,d\pi_\pm\,d\pi_\pm
-
C\int_{M_5}A_\pm\,dA_\pm\,dA_\pm,
\label{SII5-chiral}
\end{align}
up to boundary terms absorbed into \(S_{II,4}\). \footnote{Here 
$C=1/(6\sqrt{3\alpha})$. It is dropped below as it can be absorbed into field redefinitions.}

The key point is that each chiral sector now depends on a single fluid
variable, as in the Weyl case. One may therefore consider
independent admissible diffeomorphisms
\begin{align}
&i) \qquad\delta_\xi \pi_+={\cal L}_\xi \pi_+,\qquad \delta_\xi\pi_-=0,
\nonumber\\
  &ii) \qquad \delta_\eta \pi_+=0,\qquad\quad\
    \delta_\eta\pi_-={\cal L}_\eta \pi_-,
\label{chir-diff}
\end{align}
with \(\xi\) and \(\eta\) tangent to the boundary. By the Weyl
analysis, the variations of \(S_+\) and \(S_-\) reduce to vanishing
boundary terms, so the equations of motion again come from the
four-dimensional action only. 
In chiral variables, this takes the form
\begin{align}
S_{II,4}
=&
\int_{{\cal M}_4} P(\pi_+-A_+,\pi_- -A_- )
+  A_+ (\pi_+ -A_+ )d(\pi_+ +A_+ )
-  A_- (\pi_- -A_- )d(\pi_- +A_- )
\nonumber\\
&+
\int_{{\cal M}_4}(\pi_+ -A_+ )(\pi_- -A_-)
\,d\!\left[\sigma (\pi_+ -\pi_-)+\tau (A_+-A_-)\right],
\label{S-II-chir}
\end{align}
where the last term is the gauge-invariant mixing polynomial, written
in a convenient chiral form.
The parameters $(\s,\tau)$ are linearly related to $(\b,\g)$ of 
$Q_{\b,\g}$ in \eqref{Q-pol} (see Appendix~\ref{app:2fluid}).

The equations obtained from the variations \eqref{chir-diff}
have the same form as in the Weyl case,
\begin{align}
{\cal J}^\mu_\pm\partial_{[\mu}\pi_{\alpha],\pm}=0,
\qquad
\partial_\mu {\cal J}^\mu_\pm=0,
\qquad
d\pi_\pm\,d\pi_\pm=0.
\label{chir-eom}
\end{align}
Equivalently, in vector/axial variables they read
\begin{align}
d\pi\,d\tilde\pi=0,
\qquad
d\pi\,d\pi+3\alpha\,d\tilde\pi\,d\tilde\pi=0,
\label{vecax-constraints}
\end{align}
together with
\begin{align}
\partial_\mu{\cal J}^\mu=0,
\qquad
\partial_\mu\tilde{\cal J}^\mu=0.
\label{vecax-conservation}
\end{align}
These are fully consistent with the gauge equations
\eqref{J-part-cons}. The $(\s,\tau)$ coupling term in \eqref{S-II-chir} affects the explicit constitutive form of the
currents, but not the structure of the equations themselves.

We conclude that a local four-dimensional variational principle exists
for the two-fluid action when the admissible diffeomorphisms are taken
to act independently on the two chiral sectors. By contrast, common
diffeomorphisms are too weak, while naive independent variations of
\(\pi\) and \(\tilde\pi\) do not close on four-dimensional equations.
In this sense, the chiral decomposition is the natural formulation of
the two-fluid variational problem. Further explicit expressions for the
two-fluid theory are collected in Appendix~\ref{app:2fluid}.

\subsection{Equations of motion for the single-fluid action}

We finally turn to the single-fluid action. As discussed earlier, it
may be obtained from the two-fluid action by the reduction
\begin{align}
    \tilde\pi \to d\psi,
\end{align}
so that the axial one-form is replaced by the gradient of a
pseudo-scalar field. This reduction leaves the anomaly equations
unchanged. The action then takes the form
\begin{align}
\label{S-single}
S_I
=&
\int_{{\cal M}_4}P(\pi-A)
+\tilde A(\pi-A)d(\pi+A)
+\psi\bigl(d\pi\,d\pi+\alpha\, d\tilde A\,d\tilde A\bigr)
+Q_{\beta,\gamma}(\pi,A;d\psi,\tilde A)
\nonumber\\
&\qquad
-\int_{{\cal M}_5}\tilde A\,dA\,dA
+\alpha\,\tilde A\,d\tilde A\,d\tilde A\,.
\end{align}

\paragraph{General variation.}
Proceeding as before, we define the covariant currents by the full
variation of the action,
\begin{align}
  \delta S_{eff} &= \int_{{\cal M}_4}
   -\delta\pi_\mu\mathcal{J}^\mu +\delta A_\mu J^\mu
  +\delta\tilde A_\mu \tilde J^\mu +
  \delta\psi\,(d\pi d\pi+\alpha\, d\tilde A d\tilde A) 
 \\
&- \int_{{\cal M}_5} 2\delta A\, d\tilde A dA +
    \delta\tilde A\,(dA dA +3\alpha\, d\tilde A d\tilde A).
    \nonumber
\end{align}
 
 We shall again work with the covariant
currents; explicit formulas are given in
Appendix~\ref{app:1fluid}.

\paragraph{Gauge variations.}
The dynamical fields are now \(\pi_\mu\) and \(\psi\). Their admissible
gauge variations are
\begin{align}
\delta_\lambda \pi=d\lambda,
\qquad
\delta_\lambda A=0,
\qquad
\delta_{\tilde\lambda}\psi=\tilde\lambda,
\qquad
\delta_{\tilde\lambda}\tilde A=0.
\end{align}
The corresponding equations of motion are
\begin{align}
    \partial_\mu{\mathcal J}^\mu =0,
 \label{dJ-500}\\
    d\pi\,d\pi +\alpha\,d\tilde A\,d\tilde A =0.
 \label{constr-500}
\end{align}
The first is the continuity equation for the particle current, while
the second is the constraint enforced by the Wess--Zumino field. Note
that the equation obtained from varying \(\psi\) is the same whether
one regards \(\psi\) as a gauge degree of freedom or as an independent
field.

\paragraph{Diffeomorphism variations.}
We now consider diffeomorphisms acting on both dynamical variables,
\begin{align}
\delta_\xi\pi={\cal L}_\xi\pi,
\qquad
\delta_\xi\psi=\xi^\alpha\partial_\alpha\psi.
\end{align}
Since the five-dimensional part of the action depends only on the
background fields, its variation does not contribute here. The
four-dimensional variation gives
\begin{align}
{\cal J}^\mu \partial_{[\mu}\pi_{\nu]}
+\pi_\nu\,\partial_\mu {\cal J}^\mu
+\partial_\nu\psi\,
\bigl(d\pi\,d\pi+\alpha\,d\tilde A\,d\tilde A\bigr)
=0.
\end{align}
Using the gauge equations \eqref{dJ-500} and \eqref{constr-500}, the
second and third terms vanish, and one is left with
\begin{align}
{\cal J}^\mu \partial_{[\mu}\pi_{\nu]}=0.
\label{Lich-II}
\end{align}
This is again the Carter--Lichnerowicz equation for the fluid
momentum.

\paragraph{Two regimes.}
Equation \eqref{Lich-II} implies that if \({\cal J}\neq 0\), then the
two-form \(d\pi\) has degenerate rank and therefore
\begin{align}
d\pi\,d\pi=0.
\end{align}
On the other hand, if \(d\pi\,d\pi\neq 0\), then \({\cal J}=0\). By
\eqref{constr-500}, the two possibilities are controlled by the
background quantity \(d\tilde A\,d\tilde A\) (clearly assuming \(\alpha\neq 0\)).

\begin{enumerate}
\item
If \(d\tilde A\,d\tilde A=0\), then \eqref{constr-500} implies
\(d\pi\,d\pi=0\), and the equations of motion reduce to
\begin{align}
\partial_\mu \hat{\mathcal J}^\mu =0,
\qquad\qquad
\hat{\mathcal J}^\mu \partial_{[\mu}\pi_{\nu]}=0,
\end{align}
where
\begin{align}
\hat{\cal J}^\mu={\cal J}^\mu+2[d\psi\,d\pi]^\mu .
\end{align}
Thus the Wess--Zumino field effectively decouples, and one recovers the
earlier anomalous Euler equations without an independent
pseudo-scalar.\footnote{One may use the identity
\(2[d\psi\,d\pi]^\mu\partial_{[\mu}\pi_{\nu]}
=\partial_\nu\psi[d\pi\,d\pi]\).}

\item
If \(d\tilde A\,d\tilde A\neq 0\), then \eqref{constr-500} implies
\(d\pi\,d\pi\neq 0\), and the previous equations reduce to
\begin{align}
{\mathcal J}^\mu =0,
\qquad\qquad
d\pi\,d\pi +\alpha\,d\tilde A\,d\tilde A =0.
\end{align}
These are the equations obtained in our earlier work  from 
unrestricted variations of \(\pi\) and \(\psi\) (see \cite{abanov2024hydrodynamics}, Section 4.2.1). 
In this case, the condition
\({\cal J}=0\) is not unphysical, since the conserved current is
carried by
\begin{align}
\hat{\cal J}^\mu=2[d\psi\,d\pi]^\mu ,
\end{align}
which is a first integral of \(\partial_\mu\hat{\cal J}^\mu=0\),
as explained in \cite{abanov2024hydrodynamics}. 
\end{enumerate}

We conclude that the single-fluid theory again admits a local
four-dimensional variational principle based on admissible gauge and
diffeomorphism variations. We refer to our earlier work \cite{abanov2024hydrodynamics} for
a complete analysis of this theory, including the
stress-tensor equation analogous to \eqref{TW-force-500}. 

\subsection{Relation between the two-fluid and single-fluid
  theories}

In the following, we summarize and interpret the properties of these
two theories in light of the results for their respective
variations. In the case of the two-fluid action, local
four-dimensional equations of motion were obtained by passing to
chiral variables \(\pi_\pm\) and allowing independent diffeomorphism
variations of these two sectors. In that form, the equations of
motion have the same structure as in the Weyl case, although the two
chiral sectors remain coupled through the pressure term and, in
general, through the gauge-invariant mixing polynomial in the
 action \eqref{S-II-chir}.

Other possible variations did not lead to satisfactory results: common
diffeomorphism variations of \((\pi,\tilde\pi)\) are too weak to
determine two independent fluid momenta, while their independent
variations imply unacceptable equations of motion.

The natural reduction of degrees of freedom is obtained by
imposing
\begin{align}
  d\tilde\pi=0,
  \qquad\Longrightarrow\qquad
  \tilde\pi=d\psi,
\end{align}
so that the axial one-form has no independent vorticity and is
represented by the pseudo-scalar field \(\psi\). Equivalently, in
terms of the chiral variables, this condition enforces equal
vorticities in the two sectors,
\begin{align}
  d\pi_+=d\pi_-.
\end{align}
With this reduction, the two-fluid action collapses to the
single-fluid action, and the variational problem changes accordingly
to that of the fields \(\pi\) and \(\psi\).
Note that the passage from the two-fluid theory to the
single-fluid theory involves both a constraint on fields and
the variational problem specific for this subspace.
In Appendix C,  the relation between these
two theories is further discussed in a slightly more general case, 
allowing the pressure in the single-fluid theory \eqref{S-single}
to depend on both variables, 
$P(\pi-A, d\psi-\tilde A)$, as in the two-fluid case.

Summarizing, the two-fluid
theory describes interacting left and right Weyl fluids, having independent chiral variations. 
The single-fluid theory also involves a remnant axial degree of freedom
$\psi$, which does not allow independent
vorticities.

\subsection{Hydrodynamic regime as an infrared reduction}

In the following, we return to the discussion at the beginning of this Section concerning the relation between effective field theory and hydrodynamics. In the path-integral approach developed in this paper, it is clear that this correspondence is not straightforward. In the following, we propose a possible strategy for establishing it and, at the same time, for justifying the use of restricted variations of the effective fluid action. Our remarks are necessarily somewhat speculative, but we believe it is worthwhile to state them explicitly.

One of the basic assumptions of hydrodynamics is that, at low energy, fluids relax to the subspace of motions determined solely by symmetries \cite{Liu:2018kfw}. Accordingly, admissible variations should remain within
this subspace and should therefore themselves be generated by fluid symmetries \cite{carter1988standard,arnold2008topological,AbanovII}.

It follows that the low-energy effective field theories obtained above do not yet directly describe hydrodynamic regimes. In general, they contain both slow and relatively fast collective modes, determined by symmetry and dynamics, respectively. Hydrodynamics, by contrast, retains only the slow motions that survive equilibration. In this sense, passing from the effective field-theory description to hydrodynamics requires an additional reduction, or projection, which we now outline.

We assume that the effective theory enters a fluid regime in
which:
\begin{itemize}
    \item local conservation laws give rise to slow collective modes;
    \item all other non-hydrodynamic modes are fast, in the sense that
    they remain gapped or relax on a much shorter time scale;
    \item the long-wavelength dynamics is well described by local
    quasi-equilibrium.
\end{itemize}
Under these assumptions, one expects that the relevant low-energy
configurations of the fluid momentum \(\pi\) do not occupy the full
space of Abelian gauge fields, but instead lie on a reduced
``hydrodynamic'' manifold, which we denote by \({\cal F}\). We then
write
\begin{align}
    \pi = \pi_{\parallel} + \pi_\perp,
    \label{eq:pi-split-app}
\end{align}
where \(\pi_{\parallel}\in{\cal F}\) denotes the slow hydrodynamic
component, while \(\pi_\perp\) represents the fast
non-hydrodynamic component transverse to \({\cal F}\). This
decomposition is necessarily schematic, since no explicit
construction of the hydrodynamic manifold is presented here.

\paragraph{Integrating out the transverse sector.}

The effective theories described earlier, for example \eqref{S-weyl2},
may now be regarded as functionals of both slow and fast variables,
\begin{align}
    S_{\rm eff}[\pi]
    =
    S_{\rm eff}[\pi_\parallel,\pi_\perp].
\end{align}
The hydrodynamic approximation amounts to assuming that the transverse
equation can be solved perturbatively in derivatives,
\begin{align}
    \pi_\perp
    =
    \Pi_\perp[\pi_\parallel,\partial \pi_\parallel,\ldots],
    \label{eq:solve-perp-app}
\end{align}
and that this solution admits a local derivative expansion in the
infrared. Substituting \eqref{eq:solve-perp-app} back into the action
yields a reduced hydrodynamic action
\begin{align}
    S_{\rm hydro}[\pi_\parallel]
    =
    S_{\rm eff}\big[\pi_\parallel,\Pi_\perp[\pi_\parallel]\big].
    \label{eq:Shydro-reduced-app}
\end{align}

To leading order in the infrared derivative expansion one may have
\(\pi_\perp=0\), but in general the transverse sector does not simply
vanish: it contributes constitutive corrections. Therefore,
integrating out the transverse modes should not be identified too
naively with setting them equal to zero. The main assumption is that
the reduced action \eqref{eq:Shydro-reduced-app} is local and can be
interpreted as the hydrodynamic effective action.

\paragraph{Admissible variations on the hydrodynamic manifold.}

Once the reduced action \eqref{eq:Shydro-reduced-app} is obtained, its
variational problem should be defined within the slow manifold
\({\cal F}\). In other words, only tangent variations
\(\delta\pi_\parallel\) are admissible in the reduced hydrodynamic
theory:
\begin{align}
    \frac{\delta S_{\rm hydro}[\pi_\parallel]}{\delta \pi_\parallel} = 0.
\end{align}
A natural expectation is that these tangent variations are generated by
the geometric operations familiar from fluid mechanics, namely gauge
variations
\begin{align}
    \delta_\lambda \pi=d\lambda,
\end{align}
which encode the continuity equation, and diffeomorphisms
\begin{align}
    \delta_\xi \pi={\cal L}_\xi \pi,
\end{align}
which encode the Euler equation and momentum balance. The idea is that
these variations preserve the hydrodynamic sector and therefore remain
tangent to \({\cal F}\). This viewpoint is attractive because it
matches the known variational principles of Euler hydrodynamics
\cite{mobbs1982variational,carter1988standard,AbanovII,schutz1977variational}.

We conclude that the hydrodynamic equations arise by restricting the
effective action to the slow hydrodynamic sector and then varying
within it.  This argument provides a possible bridge between effective
field theory and hydrodynamics.

Most likely, the reduction can be explicitly understood in the
  real-time Schwinger--Keldysh
setting, which allows to include temperature and dissipation
\cite{Liu:2018kfw,Jensen:2018hse,RattSK}. These provide the
mechanism for relaxation of non-conserved modes, while the conserved slow
sector survives at late times and form the perfect fluid described
here.

\paragraph{Hydrodynamics and bosonization in two and four dimensions.}

The previous discussion highlights an important difference between two
and four spacetime dimensions. As seen earlier,
the restricted variations imply the
dynamics-independent constraints \(d\pi=0\) in two dimensions and
\(d\pi\,d\pi=0\) in four dimensions.
As a matter of fact, in two dimensions admissible and unconstrained variations
lead to the same equations of motion
\cite{abanov2024hydrodynamics} (see Sections~3.1, 3.2).
Solving $d\pi=0$ in terms of the scalar field, $\pi=d\theta$, shows that
the two-dimensional version of the single-fluid theory \eqref{Seff-50}
reduces precisely to the scalar field theory for 
bosonization of fermions. In this case, the relation between
hydrodynamics and effective field theory is actually an equivalence.
Therefore, the previous discussion strictly applies to higher dimensions.

Bosonization above two dimensions is expected to be a map between
interacting fermionic theories and interacting bosonic theories in their
common infrared limit.
The three-dimensional dualities provide several examples of this map
\cite{Seiberg-duality,CappelliVillaBosDual2025}.
Furthermore, the fermion-boson map has been established in
any dimension in the extreme infrared, topological limit of
the respective theories
\cite{gaiottokapustinspinTQFT1,kapustinthorngren2017,Kapustin-bosonI}.

The four-dimensional theories obtained in this work may be
useful not only for hydrodynamics but also as effective field theories for
bosonization. This is a rather interesting and
active area of research \cite{shao,fidkowski}.
The study of these theories beyond classical restricted variations,
up to general variations and then quantization,
 is rather problematic, due to  nonlinearities in particular.
The correct treatment of mixed 4d-5d actions should also be understood. 
 These issues will be addressed in future investigations.

\section{Conclusions}

In this work, we obtained three effective actions for four-dimensional
fermions by analyzing the path integral in the infrared limit,
assuming an RG flow into a fluid phase. These actions are written in
terms of the bosonic variables of fluid momenta, and correctly reproduce
the chiral anomalies for vector and axial Abelian
backgrounds. They describe, respectively, the Weyl fluid, the
single-fluid theory of the Dirac fermion, and the two-fluid
theory with independent vector and axial fluid degrees of freedom.

The form of these actions is dictated by gauge invariance and involves
topological terms corresponding to transgression forms. These are
generalizations of the Chern--Simons action for ``twin''
gauge fields, one fluid momentum and one background, with
simultaneous gauge transformations. In general,
the resulting actions are mixed 4d–5d functionals.

This derivation confirms the earlier proposals based on anomaly inflow
\cite{abanov2024hydrodynamics}, while extending them in the Dirac case
by an additional gauge-invariant term with two free dimensionless
couplings.

In the last part of this work, the fluid equations of motion were
obtained in each theory by varying the actions under gauge
transformations and diffeomorphisms. This restricted variational
principle, well known in non-anomalous Euler hydrodynamics, was
justified on two grounds. From the technical point of view, it allows
one to derive local four-dimensional fluid equations from the
five-dimensional functionals. From the physical point of view, it
respects the reduced class of fluid motions that remain after
equilibration, which is expected in hydrodynamics but must be imposed
in effective field theory.

The results of this work may have interesting developments in two
directions, namely hydrodynamics and bosonization.

In the first direction, the path-integral argument may also suggest how
temperature and entropy can be incorporated into the fluid
actions. Furthermore, the study of currents in stationary backgrounds
could lead to rather general expressions for transport coefficients,
for example those governing chiral magnetic and vortical effects. In
this context, an interesting question is whether these coefficients
depend not only on the anomalies, but also on the additional
gauge-invariant terms appearing in the actions. Another
direct extension of the present analysis would be to include the mixed
axial-gravitational anomaly in the actions
(cf.~\cite{abanov2024hydrodynamics}) and to study the corresponding
contributions to transport.  We also remark that the 
fermion path-integral argument can be repeated in three dimensions and may
suggest how to include the corresponding global parity anomaly in
hydrodynamics.

Regarding bosonization in four-dimensions, we have already pointed out
that the effective actions studied here are natural candidates for
anomalous bosonic theories. Further progress will likely require a
better understanding of the physics of mixed 4d–5d systems.

\paragraph{Acknowledgments}

We thank P. Wiegmann for many scientific exchanges on the themes of
this work. We have benefited from fruitful discussions with
N. Nekrasov and K. Jensen.

A.G.A thanks the G.~Galilei Institute for Theoretical Physics for
hospitality. His work has been supported by the National Science
Foundation under Grant NSF DMR-2116767 and by NSF-BSF grant
2022110. A.C thanks the Simons Center for Geometry and Physics for
hospitality. His work was partially supported by the grant PRIN
2022SJCKAH of the Italian Ministry of University and Research. The authors thank the organizers of the Emergo26 workshop at MPI-PKS in Dresden for their hospitality, where this work was completed.

\appendix


\section{General form of transgressions}
\label{app:trans}

In this Appendix, we present the general derivation of (Abelian)
transgression forms, paying attention to the free parameters that can
arise.

We start from integrating the difference of Chern classes,
$F^3_\pi-F^3_A$, leading to the transgression for
the $AdAdA$ anomaly term. We first use the standard linear interpolation
 method, as follows:
\begin{align}
 &   A_t=t\pi +(1-t) A, \qquad\qquad t\in [0,1],
 \nonumber \\
&    F^3_\pi-F^3_A=\int_0^1 dt\, \d_t \left( F_t^3 \right) =
                  3d \int_0^1 dt (\pi -A) F^2_t,
 \nonumber \\
&    T_5(\pi,A)=3 \int_0^1 dt (\pi -A) F^2_t,
 \label{T5-int-100}
\end{align}
where we used $\d_t F_t =d\d_t A_t =d(\pi-A) $. The integral
expression \eqref{T5-int-100} shows that the transgression is gauge
invariant by construction, because $\pi-A$ and $F_t$ are invariant.

Integration of \eqref{T5-int-100} gives the expression
\begin{align}
    T_5(\pi,A) &= \pi d\pi d\pi -A dA dA +d\left[A\pi d(\pi+A)\right]
 \label{T5-500}\\
    &= C_5(\pi)-C_5(A) +dB_4(\pi,A)\,.
 \nonumber
\end{align}
reported in the main text.\footnote{In the main text we write \(A(\pi-A)d(\pi+A)\), which is
equivalent to the present form because \(A\wedge A=0\).}

Next, we ask ourselves whether this result is unique or, rather, other
interpolations may give different expressions. We consider the general case
\be
A_t=f(t) \pi +g(t) A, \qquad {\it with}\qquad
f(0)=0,\ f(1)=1,\ g(0)=1,\ g(1)=0, 
\ee
The expression
\be
F_\pi^3-F_A^3= 3 d\int_0^1 dt \frac{d}{dt}\left(f\pi+g A\right)
\left(d A_t\right)^2\,,
\ee
is gauge invariant provided that $f'=-g'$, whose solution is $g=1-f$
for the given boundary conditions. Thus, we rewrite
\be
  \label{AAA-int}
  F_\pi^3-F_A^3= 3d\int_0^1 dt f'(\pi- A) \left(d(f\pi +(1-f) A)\right)^2
=3 d\int_0^1 df (\pi- A) \left(d(f\pi +(1-f) A)\right)^2 .
\ee
This shows that the form \eqref{T5AAA-100} of the transgression
$T^{AAA}$ is uniquely determined, as well as the hydrodynamic action
$S_W$ \eqref{S-weyl} for the Weyl fermion fluid.

Next, the transgression for the $\tilde AdAdA $ anomaly is obtained by
interpolating  $F_{\tilde \pi} F^2_\pi $ and $F_{\tilde A} F^2_A$.
For linear interpolation, $A_t = t \pi+(1-t) A$ and 
$\tilde A_t = t\tilde \pi +(1-t) \tilde A$, we write
\begin{align}
  F_{\tilde\pi}F^2_\pi- F_{\tilde A} F^2_A &=
dT_5^{AVV} (\pi,A;\tilde\pi, \tilde A)=
 \int_0^1 dt \frac{d}{d t} \left(d\tilde A_t dA_t dA_t \right)
 \nonumber \\
&= d \int_0^1 dt\,\Big[(\tilde\pi-\tilde A)dA_tdA_t
                +2(\pi-A)d\tilde A_t dA_t],
\label{AVV-int}
\end{align}
Upon integration over $t$ we obtain, after some calculations,
\begin{align}
    T_5^{AVV} (\pi,A;\tilde\pi, \tilde A)
    &= \tilde \pi d\pi d\pi -\tilde A dA dA 
    + d\Big[\tilde A(\pi-A)d(\pi+A)\Big] \nl
    & + \frac{1}{3}d\Big[(\tilde \pi-\tilde A)(\pi-A) d(2\pi+A)\Big]\,.
 \label{T5AVV-200}
\end{align}

The expression in the second line corresponds to a non-minimal gauge-invariant
4d term, which respects all the defining properties of
the transgression. It clearly has a two-parameter generalization,
given by $dQ_{\b,\g}(\pi,A;\tilde\pi, \tilde A)$ \eqref{Q-pol} in the main text.
The above expression corresponds to $(\b,\g)=(2/3,1)$.

The origin of this two-parameter freedom is understood by repeating the
integration for general interpolations  $A_t=f(t) \pi +g(t) A$ and
$ \tilde A_t =\tilde f(t) \tilde \pi +\tilde g(t) \tilde A $, where
again $g=1-f$ and $\tilde g =1-\tilde f$ for gauge invariance.
We obtain the expression 
\begin{align}
  \label{TAVV-300}
  &  T_5^{AVV}(\pi,A;\tilde \pi, \tilde A)=
\\
&= \int_0^1\!\!\! dt \tilde f'(\tilde\pi -\tilde A)
  \left(d(f\pi +(1-f) A)\right)^2 +
  2f'(\pi-A)d(f\pi+(1-f)A)d(\tilde f \tilde \pi +(1-\tilde f)\tilde A)
\nl
 &= \int_0^1 dt \tilde f'(\tilde\pi -\tilde A)
  \left(d(t\pi +(1-t) A)\right)^2 +
   2(\pi-A)d(t\pi+(1-t)A)d(\tilde f \tilde \pi +(1-\tilde f)\tilde A).
\nonumber
\end{align}
After replacing again $dt f' \to dt$ in the third line above, one is left
with an expression involving the numerical integrals
$\int \tilde f' =1$, $\int \tilde f' t=1-\int \tilde f$ and
$\int \tilde f' t^2 =1 - 2\int \tilde f t$, which corresponds to two
undetermined constants. We thus proved that 
$T_5^{AVV}$ admits a gauge invariant polynomial with two free parameters.

Another way to understand this freedom is to consider the Dirac
anomaly in the chiral-split basis, as described in Section 5.2.
The Chern--Simons actions AAA and AVV split into single-field
expressions, $C_5(A_+)$ and $C_5(A_-)$, which have unique transgressions,
as described at the beginning of this Appendix.
The additional term must couple the two
chiralities $(\pi_+,A_+)$ and $(\pi_-,A_-)$, must be fourth-order,
vanish when one of them is set to zero, and be gauge and parity invariant.
Its form is easily guessed: it is given in \eqref{S-II-chir}, and 
involves two free parameters.

In conclusion, the general hydrodynamic actions for one fluid
\eqref{Seff-psi-200} and two fluids \eqref{Seff-II-100}
contain a two-parameter gauge invariant term.


\section{Detailed analysis of the two-fluid theory}
\label{app:2fluid}

In this Appendix we derive explicit expressions for the vector and
axial currents obtained from the two-fluid action \eqref{Seff-II-100}.
Keeping the
two-parameter gauge-invariant term in its general form, we write
\begin{align}
  \label{Seff-II-app}
  S_{II} &= S_{II,4}+S_{II,5}
 \nonumber \\
  =&\int_{{\cal M}_4} P(p,\tilde p)
     + \tilde A\,p\,d(p+2A)
     + \alpha\,\tilde A\, \tilde p\, d(\tilde p+2\tilde A)
     + \int_{{\cal M}_4}
     \tilde p\,p\,
     d\bigl(\beta\,p+\gamma A\bigr)
 \nonumber\\
  &+ \int_{{\cal M}_5}\tilde \pi\, d\pi\, d\pi
       +\alpha\,\tilde \pi\, d\tilde\pi\, d\tilde\pi
       - \tilde A\, dA\, dA
       -\alpha\, \tilde A\, d\tilde A\, d\tilde A \,.
\end{align}
As usual, we introduce the gauge-invariant one-forms
\begin{align}
  p=\pi-A,
  \qquad
  \tilde p=\tilde\pi-\tilde A.
\end{align}

The consistent currents are obtained by varying only the
four-dimensional part of the action \(S_{II,4}\), while the covariant
currents additionally include the boundary terms arising from the
variation of \(S_{II,5}\), as explained in the main text.

The consistent currents, written as dual three-forms, are
\begin{align}
  J_{\rm cons}
  &=
  n
  +2\tilde A\,dA
  +p\,d\tilde A
  +\Big[\tilde p\,d(\beta p+\gamma A)
  +(\gamma-\beta)\,d(\tilde p\,p)\Big],
  \\
  \tilde J_{\rm cons}
  &=
  \tilde n
  +p\,d(p+2A)
  +2\alpha\,\tilde A\,d\tilde A
  +\alpha\,\tilde p\,d(\tilde p+3\tilde A)
  -p\,d(\beta p+\gamma A).
\end{align}
Here \(n\) and \(\tilde n\) denote the three-forms dual to the
densities
\begin{align}
  n^\mu=-\frac{\partial P}{\partial p_\mu},
  \qquad
  \tilde n^\mu=-\frac{\partial P}{\partial \tilde p_\mu},
\end{align}
that is,
\begin{align}
  n=* \left(-\frac{\partial P}{\partial p_\mu}\,dx^\mu\right),
  \qquad
  \tilde n=* \left(-\frac{\partial P}{\partial \tilde p_\mu}\,dx^\mu\right).
\end{align}

The difference between the two kinds of currents is
\begin{align}
  J &= J_{\rm cons} -2\tilde A\,dA,
 \\
    &= n
  +p\,d\tilde A
  +\Big[\tilde p\,d(\beta p+\gamma A)
  +(\gamma-\beta)\,d(\tilde p\,p)\Big],
  \\
  \tilde J &= \tilde J_{\rm cons} -2\alpha\,\tilde A\,d\tilde A
 \\
    &= \tilde n
  +p\,d(p+2A)
+\alpha\,\tilde p\,d(\tilde p+3\tilde A)
  -p\,d(\beta p+\gamma A),
\end{align}
and arises from the boundary terms in the variation of \(S_{II,5}\),
cf.~\eqref{b-t}.

The covariant hydrodynamic currents are obtained by varying the action
with respect to \(\pi\) and \(\tilde\pi\), as described in
\eqref{S-II-var}. One finds
\begin{align}
  \mathcal{J}
  &=
  J
  -2p\,d\tilde A
  -2\tilde p\,d(p+A)
  -\gamma\,d(\tilde p\,p),
  \\
    &= n
  -p\,d\tilde A -2\tilde p\,d(p+A)
  +\Big[\tilde p\,d(\beta p+\gamma A)
  -\beta\,d(\tilde p\,p)\Big],
 \\
  \tilde{\mathcal{J}}
  &=
  \tilde J
  -p\,d(p+2A)
  -3\alpha\,\tilde p\,d(\tilde p+2\tilde A),
 \\
    &=\tilde n
-\alpha\,\tilde p\,d(2\tilde p+3\tilde A)
  -p\,d(\beta p+\gamma A).
\end{align}
As a consequence,
\begin{align}
  dJ
  &=
  d\mathcal{J}
  +2\,d\pi\,d\tilde\pi
  -2\,dA\,d\tilde A,
  \\
  d\tilde J
  &=
  d\tilde{\mathcal{J}}
  +d\pi\,d\pi
  -dA\,dA
  +3\alpha\bigl(d\tilde\pi\,d\tilde\pi-d\tilde A\,d\tilde A\bigr).
\end{align}
In particular, the dependence on the gauge-invariant term, namely on
\(\beta\) and \(\gamma\), drops out of the conservation laws.

As described in the main text, the equations of motion obtained from
the gauge variations \(\pi\to \pi+d\lambda\) and
\(\tilde\pi \to \tilde\pi +d\tilde \lambda\) take the form
\eqref{J-part-cons}. Substituting them into the previous relations,
one recovers the anomaly equations \eqref{J-II-anom}.

\bigskip
\paragraph{Contribution by the two-parameter gauge-invariant term}

Using the chiral variables \eqref{chir-split}, and after several
integrations by parts, the action \eqref{Seff-II-app} can be rewritten
as
\begin{align}
  \label{Seff-II-app-split}
  S_{II} = &\int_{{\cal M}_4} P(p_+,p_-)
     + C A_+p_+ d(p_++2A_+)
     -C A_-p_-d(p_-+2A_-)
     \nonumber\\
  &\qquad
     +\frac{C}{2} \int_{{\cal M}_4}
     p_+ p_-\,
     d\!\bigl((3\beta-2)(p_++p_-)+(\gamma-3)(A_++A_-)\bigr)
     \nonumber\\
  &\qquad
     + C\int_{{\cal M}_5}\pi_+\, d\pi_+\, d\pi_+
       -\pi_-\, d\pi_-\, d\pi_-
       -  A_+\, dA_+\, dA_+
       +A_-\, d A_-\, d A_- \,.
\end{align}
Here the constant \(C=\frac{1}{6\sqrt{3\alpha}}\). In the main text it
is omitted, since it can be absorbed by an overall rescaling of the
fields or, equivalently, of the pressure.

The contribution of the gauge-invariant term with free parameters
\(\beta\) and \(\gamma\) is
\begin{align}
  \label{Spoly-app-100}
  \Delta S_{II} = \frac{C}{2}\int_{{\cal M}_4}
  p_+\,p_-\, d\Big[3\beta\,(p_++p_-)+\gamma\,(A_++A_-)\Big].
\end{align}
The variation of \eqref{Spoly-app-100} with respect to
the dynamical fields is
\begin{align}
\delta \Delta S_{II} =-\int_{{\cal M}_4}\left(\delta\pi_+\, \Delta{\cal J}_+
+\delta\pi_-\, \Delta{\cal J}_- \right),
\label{varSpoly-app-100}
\end{align}
where the shifts of hydrodynamic currents are
\begin{align}
    \Delta{\cal J}_+ &= -\frac{C}{2}p_- \,d\Big(3\beta(p_++p_-) +\gamma (A_++A_-)\Big) -\frac{3\beta C}{2}d\Big(p_+p_-\Big),
 \\
    \Delta{\cal J}_- &= \frac{C}{2}p_+ \,d\Big(3\beta(p_++p_-) +\gamma (A_++A_-)\Big) -\frac{3\beta C}{2}d\Big(p_+p_-\Big), 
\end{align}
Accordingly, the gauge-type equations of motion become
\begin{align}
d({\cal J}_++\Delta {\cal J}_+)+3C\,d\pi_+\,d\pi_+=0,
\label{gaugeEOMRpoly-app-100}
\\
d({\cal J}_-+\Delta {\cal J}_-)-3C\,d\pi_-\,d\pi_-=0.
\label{gaugeEOMLpoly-app-100}
\end{align}
Similarly, the diffeomorphism equations retain the same form, with the
currents shifted as
\begin{align}
{\cal J}_+\to {\cal J}_+ +\Delta {\cal J}_+,
\qquad
{\cal J}_-\to {\cal J}_- +\Delta {\cal J}_-.
\label{Jshift-poly-app-100}
\end{align}

Since \eqref{Spoly-app-100} is manifestly gauge invariant, it does not
modify the anomaly polynomial and therefore leaves the anomaly
equations unchanged. What changes are the explicit expressions for the
hydrodynamic currents, which are shifted by local gauge-invariant
terms. Thus the gauge and diffeomorphism Ward identities keep the same
form, provided they are written in terms of the corrected currents.

\section{Detailed analysis of the single-fluid theory}
\label{app:1fluid}

Let us now consider the single-fluid defined by the action
\begin{align}
S_I[\pi,\psi]=&
\int_{{\cal M}_4} P(p,\tilde p) + \tilde A\, p\,d(p+2A)+
\psi (d\pi d \pi +\a d \tilde A d\tilde A)
\nl
 &+\int_{{\cal M}_4} \tilde p\, p\,d(\beta p +\gamma A)
 -\int_{{\cal M}_5} \tilde AdAdA+ \a \tilde A d\tilde A d\tilde A.
  \label{Seff-500}    
\end{align}
Here we added the dependence of pressure on $\tilde p =d\psi-\tilde A$ omitted in the main text (c.f. Eq.~\eqref{S-single}). This will be useful for
comparing with the two-fluid theory in the previous Appendix.

We compute the general variation of this action and obtain
\begin{align}
  \delta S_{eff} &= \int_{{\cal M}_4}
   -\delta\pi_\mu\mathcal{J}^\mu +\delta A_\mu J^\mu
  +\delta\tilde A_\mu \tilde J^\mu +
  \delta\psi\,[d\tilde{\cal J} +d\pi d\pi ] 
 \label{gvar-1fluid}\\
&- \int_{{\cal M}_5} 2\delta A\, d\tilde A dA +
    \delta\tilde A\,(dA dA +3\alpha\, d\tilde A d\tilde A).
    \nonumber
\end{align}
Here we introduced $n=-\p P/\p p$, $\tilde n =-\p P/\p \tilde p$, and 
\begin{align}
    {\cal J} &= n -2\tilde p\, d(p+A) -p\,d\tilde A +\tilde p\,d(2\beta p+\gamma A) - \beta p\,d\tilde p,
 \\
    J &= n +p\,d\tilde A +\tilde p\,d(2\beta p +\gamma A) -\beta p\,d\tilde p +\gamma\,d(\tilde p p),
 \\
    \tilde{\cal J} &= \tilde n -\alpha \tilde p\,d(2\tilde p+3\tilde A) -p\,d(\beta p +\gamma A)
 \label{1f-calJ} \\
    &= \tilde n -\alpha \tilde p\, d\tilde A -p\,d(\beta p +\gamma A),
 \nonumber \\
    \tilde J &= \tilde n +p\,d(p+2A) +\alpha\tilde p\,d(\tilde p +3\tilde A) - p\,d(\beta p +\gamma A)
 \nonumber \\
    &= \tilde n +p\,d(p+2A) +2\alpha\tilde p\,d\tilde A - p\,d(\beta p +\gamma A).
\end{align}
It is important that the variation \eqref{gvar-1fluid} defines $\tilde{\cal J}$ only up to the exact form. We used this freedom to made a particular choice so that  \eqref{1f-calJ} coincides with the expression for the two-fluid theory. All expressions for currents coincide with the two-fluid expressions upon reduction $\tilde\pi \to d\psi$.

We obtain the relations
\begin{align}
    dJ &= d{\mathcal J} -2\,dA d\tilde A,
 \\
    d\tilde J &=  d\tilde{\mathcal J} +d\pi d\pi -dA dA -3\alpha\, d\tilde A d\tilde A,
\end{align}
which fit the result of gauge variations.

The equations of motion obtained from gauge variations $\delta\pi = d\lambda$ and $\delta \psi =\tilde \lambda$ give
\begin{align}
    d{\mathcal J} &= 0,
  \label{contcalJ}\\
    d\tilde{\mathcal  J} &= -d\pi d\pi,
 \label{conttcalJ}
\end{align}
which are equivalent to 
\begin{align}
    d J &= -2dA d\tilde A,
 \\
    d\tilde J &= -dA dA -3\alpha\,d\tilde A d\tilde A.
\end{align}
So far equations are the same as in the case of the two-fluid theory after the substitution $\tilde \pi =d\psi$. 

Let us now discuss the diffeomorphism variations over $\pi$ and $\psi$. Here the difference with the two-fluid theory is significant. The diffeomorphism variation of the pseudoscalar $\psi$ produces the continuity equation \eqref{conttcalJ}. The variation of $\pi$ gives
\begin{align}
    &\mathcal{J}^\mu(\p_\mu\pi_\nu-\p_\nu\pi_\mu) +\pi_\nu \p_\mu \mathcal{J}^\mu =0.
 \label{eom-505}
\end{align}
Multiplying by $\mathcal{J}^\nu$ we reproduce \eqref{contcalJ} and then obtain from \eqref{eom-505}
\begin{align}
    &\mathcal{J}^\mu(\p_\mu\pi_\nu-\p_\nu\pi_\mu) 
    = 0\,.
 \label{eom-506}
\end{align}
For nonvanishing $\mathcal{J}$ it follows from here that $d\pi d\pi =0$.

\bibliographystyle{JHEP}
\bibliography{hydro-anomalies.bib}

\end{document}